 \documentclass[final,5p,times,twocolumn]{elsarticle}

\usepackage{graphicx}
\usepackage{epstopdf}
\usepackage{subfigure}

\usepackage{amssymb}

\usepackage{algorithm}
\usepackage{algorithmicx}
\usepackage{algpseudocode}
\newcommand{\algorithmicbreak}{\textbf{break}}
\newcommand{\BREAK}{\State \algorithmicbreak}

\newcommand{\tabincell}[2]{\begin{tabular}{@{}#1@{}}#2\end{tabular}}

\usepackage{balance}

\journal{Computer Networks}

\begin{document}

\begin{frontmatter}



\title{Data Dissemination Using Interest-Tree in Socially Aware Networking}


\author[DUT]{Feng Xia\corref{cor}}
\author[DUT]{Qiuyuan Yang}
\author[DUT]{Jie Li}
\author[PolyU]{Jiannong Cao}
\author[DUT]{Li Liu}
\author[DUT]{Ahmedin Mohammed Ahmed}

\cortext[cor]{Corresponding author; Email: f.xia@ieee.org}

\address[DUT]{School of Software, Dalian University of Technology, Dalian 116620, China}
\address[PolyU]{Department of Computing, Hong Kong Polytechnic University, China}

\begin{abstract}
Socially aware networking (SAN) exploits social characteristics of mobile users to streamline data dissemination protocols in opportunistic environments. Existing protocols in this area utilized various social features such as user interests, social similarity, and community structure to improve the performance of data dissemination. However, the interrelationship between user interests and its impact on the efficiency of data dissemination has not been explored sufficiently. In this paper, we analyze various kinds of relationships between user interests and model them using a layer-based structure in order to form social communities in SAN paradigm. We propose Int-Tree, an Interest-Tree based scheme which uses the relationship between user interests to improve the performance of data dissemination. The core of Int-Tree is the interest-tree, a tree-based community structure that combines two social features, i.e. density of a community and social tie, to support data dissemination. The simulation results show that Int-Tree achieves higher delivery ratio, lower overhead, in comparison to two benchmark protocols, PROPHET and Epidemic routing. In addition, Int-Tree can perform with 1.36 hop counts in average, and tolerable latency in terms of buffer size, time to live (TTL) and simulation duration. Finally, Int-Tree keeps stable performance with various parameters.
\end{abstract}


\end{frontmatter}


\section{Introduction}
The popularity of mobile devices such as smart phones has increased contact opportunities between mobile users in pervasive environments. In this setting, mobile carriers (i.e, human beings) communicate via Bluetooth and Wi-Fi technologies in order to share different kinds of information (such as photos, commercial trades, etc.) between interested users. However, opposed to conventional mobile ad hoc networks, an end-to-end connectivity between users might not be exist and they have to carry messages until a new contact is established \cite{Pelusi:Opportunistic}. Delay Tolerant Networks (DTNs) \cite{Warthman:DTNs, Spyropoulos:Routing} are special kinds of networks which use \textit{store-carry-and-forward} scheme to forward data between disconnected users. In this paradigm, mobile devices mirror movement patterns and attributes of their owners (i.e., users), hence social characteristics and features of users can be exploited to improve the performance of forwarding protocols. This drives the emergence of socially aware networking (SAN) \cite{Xia:SAN, Vastardis:Multi-phase}, which aims at exploring social relationships and properties of network users to streamline routing and data dissemination protocols \cite{RoutingSurvey-Zhu, Xia-TPDS2014, Xia-TC2014}.

Social attributes, relationships, and behaviors of mobile users are relatively stable through a long period and they have long-term characteristics. Hence, they have been extensively used to improve the performance of data forwarding algorithms. Contemporary researches in this area mainly use social network analysis \cite{SNA-book} techniques to extract different social properties of users. The commonly used social properties include social similarity, tie strength, community, node centrality, mobility pattern, etc. Among these, the community structure has been widely used in recent routing protocols such as in \cite{Fan:Geocommunity, Xiao:Community-aware, Wu:Homing}. In this strategy, socially similar individuals form a community where the similarity could be friendship, common visiting locations, or similar interests. Generally, individuals inside a community contact one another frequently and this can be beneficial for choosing a proper forwarder within a community.

An additional important social factor that is used extensively to improve the efficiency of data dissemination protocols in SAN paradigm is interest. This is because people with similar interests meet each other frequently and share more data with each other \cite{McPherson:Homophily}. Hidi \cite{Hidi:Interest} provided theoretical proofs on what interest is and how it drives human beings in acquiring knowledge. Studying human behaviors \cite{Zhou:Role, Han:Modeling, Shang:Interest-Driven} also showed that daily activities of users in social networks such as browsing, cooperation in online societies and playing on-line games are mainly driven by their interests. Furthermore, existing routing protocols \cite{Chen:Leveraging, Costa:Socially-aware, Zhu:Ripple} validated the value of user interests for data dissemination. Specifically, interests can be assigned equal values of importance as keywords \cite{Chuah:Cooperative}, or be extracted from various vectors according to similarity for routing algorithm design \cite{Chen:Leveraging}. In addition, user interests can constitute interest lists \cite{Costa:Socially-aware, Zhu:Ripple} to help to predict movement patterns or build a multi-cast tree. However, all these interest-based approaches have not explored thoroughly on the inherent relationships between user interests and their effect on data dissemination.

In this paper, we get the inspiration from the above-mentioned two social features, i.e., community structure and user interests, in order to answer the following research questions: 1) which kinds of relationships can be available between user interests, 2) how to model the interrelationship between user interests, and 3) how the relationships between user interests affect the performance of a data dissemination protocol. To achieve these goals, we analyze relations between user interests and devise an \textbf{Int}erest\textbf{-Tree} based scheme (Int-Tree) for data dissemination. First, we build the \emph{interest-tree}, a tree-based community structure according to interests of users which is updated dynamically. Then, density of a community and social tie are calculated by social awareness. When a source node contacts an intermediate node, they update their density and social tie information. After that, Int-Tree decides whether the intermediate node is suitable to be a forwarder in accordance with different criteria in forwarding strategy module.

Hereby, we intend to clarify three points: 1) Int-Tree is the name of our scheme whereas interest-tree is the structure we constructed for presentation of community structure; 2) We divide communities based on user interests (i.e. one interest representing one community); and 3) We focus on the effects of relations between user interests and simplify the problem as one-to-one (i.e. one source and one destination) dissemination, which is the basis for one-to-many dissemination.

Int-Tree is based on our previous work BEEINFO \cite{Xia:BEEINFO}. The major difference between this work and the prior one lies on the exploration of relationships between user interests and how they affect the performance of data dissemination in a community-based SAN paradigm. Our major contributions to support this idea are summarized as follows:

\begin{itemize}
\item We study the relationships between user interests in SAN paradigm and propose a layered model to map user interests into different levels. Our model is able to present various elements of relationships between user interests including interest inclusion, cross-layer interests and interest intersection.
\item We build an interest-tree to illustrate interest inclusion which is a special kind of relationship between user interests. The structure can support to combine the social features density of a community and social tie for data dissemination.
\item We conduct extensive simulations that demonstrate the performance and effectiveness of Int-Tree in comparison to Epidemic \cite{Vahat:Epidemic} and PROPHET \cite{Lindgren:Probabilistic} protocols in terms of delivery ratio, overhead, average latency and hop count. Simulation duration, buffer size and Time-to-Live (TTL) are the most important parameters which are adopted in our simulation.
\item We also carry out further simulations to explore how Int-Tree performs under different values of $\gamma$, and evaporation factors ($\alpha$ and $\beta$).
\item Considering the situation of multiple interests, we provided discussions on the challenges and solutions.
\end{itemize}

The rest of the paper is organized as follows. An overview on interest-based data dissemination protocols, as well as community-based forwarding algorithms is presented in Section 2. Section 3 describes our interest-based data dissemination problem and Section 4 presents a layered model to analyze the relationships between user interests. Section 5 describes Int-Tree as well as the components. Section 6 presents the simulation results, compares the performance of Int-Tree, PROPHET and Epidemic protocols, and explores the influence of changing parameters on Int-Tree. Section 7 discusses some problems raised by multi-interest situation, alongside the solutions. Finally, Section 8 concludes this paper.

\section{Related Work}
Several well-designed data forwarding protocols have been proposed for DTNs that were inspired from Epidemic \cite{Vahat:Epidemic} and PROPHET \cite{Lindgren:Probabilistic} routing algorithms. These algorithms were mainly proposed for intermittently connected networks and did not use the social characteristics of users. In the Epidemic routing, messages are flooded to encounter nodes with unlimited replication policy which results in a data congestion problem in the network. To cope with this problem, several routing protocols have been proposed aiming at limiting the number of message replicas and leveraging a tradeoff between resource usage and message delivery. The PROPHET is a controlled flooding algorithm which makes use of delivery predictability metric to estimate the probability of next relay nodes to deliver messages to destination nodes. These protocols are the foundation of our work since we use the contact history of network nodes to predict the future contacts between the nodes.

According to a definition of interest given by Hidi \cite{Hidi:Interest}, interest has positive effects, such as contributing to increasing comprehension, and motivating thoughts and actions of people. The theory led researchers to study interest-based human behaviors. Zhou \textit{et al.} \cite{Zhou:Role, Han:Modeling, Shang:Interest-Driven} proved that some human activities such as rating movies, web browsing, and mobile phone text-messages are interest-driven. They also proposed an interest-driven model and explained the observed relationship between activities and power-law exponents. However, the proposed model did not consider which factors affect the user interests. To address this shortcoming, Yan \textit{et al.} \cite{Yan:Modeling} studied the posting behavior of micro-blog users in mobile internet and concluded that social concern affect users' interest. Additionally, they proposed a model with social concern to slow down the decay of interest. Carofiglio \textit{et al.} \cite{Carofiglio:ICP} presented Interest Control Protocol (ICP) to achieve fully efficient and fair flow control for content-centric networking. In spite of the fact that the mentioned methods have explored the user interest from social network perspective and proved that user interests affect human behaviors, they still failed in describing how interest drives human behaviors.

As for socially aware data dissemination, Costa \textit{et al.} \cite{Costa:Socially-aware} proposed SocialCast and exploited predictions based on metrics of social interaction (e.g. receivers' interests, social ties) to identify the best information carriers. In SocialCast, each node broadcasts a control message to its neighbors containing the list of interests and the corresponding list of utility values. This information is key to making message forwarding decisions. Similarly, Zhu \textit{et al.} \cite{Zhu:Ripple} applied interest list to their publish/subscribe service, named Ripple, in order to update request dissemination in the cloud.

In \cite{Chuah:Cooperative}, the authors introduced Cooperative User Centric Information Dissemination (CUCID) Scheme. The scheme uses keyword space to describe user interests and each user has a vector of probabilities describing how interested the user is in data items described by the keywords. Chen \textit{et al.} \cite{Chen:Leveraging} proposed SPOON that used an interest extraction algorithm to derive a node's interests, file vector, group vector and node vector. Moreover, it groups common-interest nodes into communities by calculating their similarities. Bjurefors \textit{et al.} \cite{Bjurefors:Interest} examined the data-centric architecture of Haggle and explored how interests and data were presented and distributed. Specifically, nodes express their interests in the form of attributes. A node description is a data object that consists of the node's interests and metadata. The authors in \cite{Gao:User} considered contact patterns and interests of users to ensure effective data relaying.

The discussed algorithms in the last two paragraphs explored the impact of user interests on data dissemination. Ripple and Haggle studied the overlap of interests, while SPOON used similarity of vectors to reflect an indirect relation between user interests. In SocialCast, CUCID, and \cite{Gao:User}, the relation between user interests have not been exploited.

Dissimilar to the above efforts, Wu and Wang \cite{Wu:Hypercube} recently extracted nodes' social features to help accomplish the multipath routing. They believe that \emph{people contact with each other more frequently if they have more social features in common}, which is nearly the same as what we stated in this paper. However, our work differs from theirs by focusing on the relationships of user interests and the consequent effects.

Beyond interest and social ties, community is also a widely used concept for data forwarding in SAN paradigm. Some community-based protocols have been proposed. In LABEL \cite{Hui:How}, nodes deliver messages only to the members in destination community. BUBBLE RAP \cite{Hui:Bubblerap} uses community and central nodes to construct socio-aware overlay for effective routing. The main drawback of these methods is high cost of constructing and maintaining the network overlay. LocalCom \cite{Li:Localcom} and Gently \cite{Musolesi:Writing} take inter-community and intra-community routings into consideration. LocalCom adopts similarity feature for further detection of community, while Gently uses LABEL and CAR (Context Adaptive-aware Adaptive) \cite{Musolesi:Adaptive} for different routing phases. Home spread (HS) \cite{Wu:Homing} considers the locations as \textit{community homes} or \textit{homes} where nodes visit frequently. The concept of \textit{community homes} is also adopted in Community-Aware Opportunistic Routing (CAOR) \cite{Xiao:Community-aware}. Another example is BEEINFO \cite{Xia:BEEINFO} in which the authors introduced the concept of community density and divided communities according to users interests. In this method, optimal relay nodes are chosen based on community density and social tie.

In this paper, we go one step further by exploring the relationship between user interests and its impact on data dissemination in SAN paradigm. To this end, a tree-based structure is presented in order to model user interests and their relationships. Based on the tree structure, the network is divided into different communities, combining community density and social tie to support data dissemination. Our method differs from the existing protocols by considering the fact that the interrelation between user interests are used to form a community structure which improves the performance of data dissemination in SAN paradigm.

\section{Problem Statement}
In most of the interest-based data dissemination protocols, it is commonly assumed that users have a list of interests and they give little attention to the relations between user interests. To address this issue, we present a real life scenario to highlight the problem to be solved as follows:

\begin{figure}[!ht]
\centering
\includegraphics [width=3.5in]{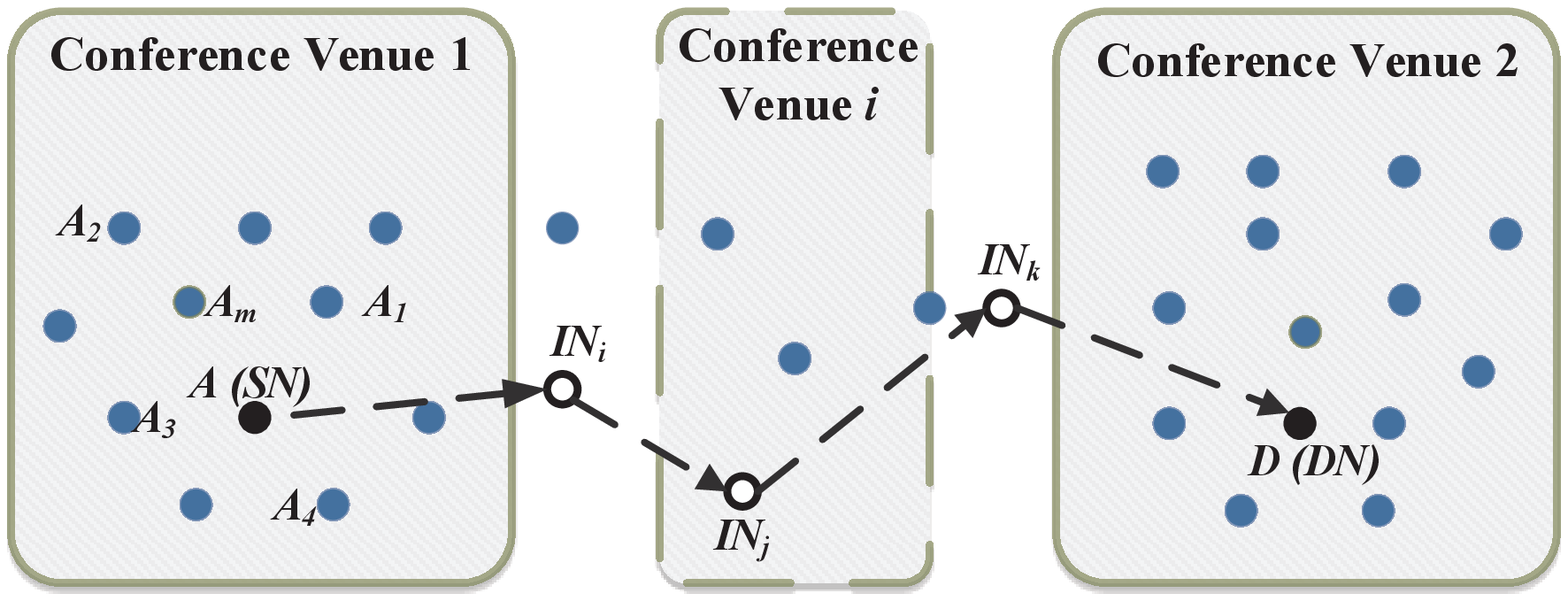}
\caption{A real life scenario where Alice ($A$) has some items to forward to David ($D$). The empty circles and the dashed arrows constitute a possible route from $A$ to $D$.}
\label{scenario}
\end{figure}

Suppose that Alice ($A$) is attending a presentation during an academic conference at Conference Venue 1, along with other attendees ($\{A_{1}, A_{2}, ..., A_{m}\}$), as shown in Fig. \ref{scenario}. There are other venues in the meantime and Conference Venue $i$ is a random venue represented by dashed round rectangle. Inspired by the presentation, she finds interesting items, such as a paper related to the topic or a video clip she just recorded. She wants to share these items to a friend of hers, David ($D$). However, $A$'s movement is restricted for a while because of the presentation and $D$ is out of $A$'s range being in another venue. If the backbone network can not provide service, the direct contact between $A$ and $D$ will not be possible. Thus, to finish the dissemination, $A$ has to rely upon other attendees who can contact $D$ with higher possibility. Fig. \ref{scenario} also shows a possible route from $A$ to $D$ in this scenario. In the figure, the black dots represent $A$ and $D$. The empty circles are the potential forwarders. The route from $A$ to $D$ consists of dashed arrows, meaning that there might be more forwarders on the route. The other dots are random nodes in the network.

Here comes the problem: how to choose proper forwarders among the potentials to achieve efficient dissemination? Our solution to this problem is based on interests since people with the same interest meet with one another in higher probability. Additionally, user interests correlate to each other. Therefore, we examine the relations between user interests to help $A$ choose forwarders to the target destination.

\section{Relationship between User Interests}
In social networks, people usually have a few interests which are stable over a time. Homophily \cite{McPherson:Homophily} describes that similar characteristics between individuals generally result in a bond between them. As the saying ``birds of a feather flock together'' suggests, people with similar interest, backgrounds or beliefs tend to form stronger relationships than those with dissimilar ones. This statement can also be supported by a social network theory which is proposed in \cite{Katsaros:Social}. Based on this theory, the file interests have clear structured categories, and for the majority of users, 80\% of the shared files fall into 20\% of the file categories \cite{Fast:Creating}. This is obvious on a campus or a conference venue. For example, the attendees in the same conference room share the same interest with the speech or the keynote, whereas those in different venues are likely to be interested in dissimilar topics.

\begin{figure}[!ht]
\centering
\includegraphics [width=3.5in]{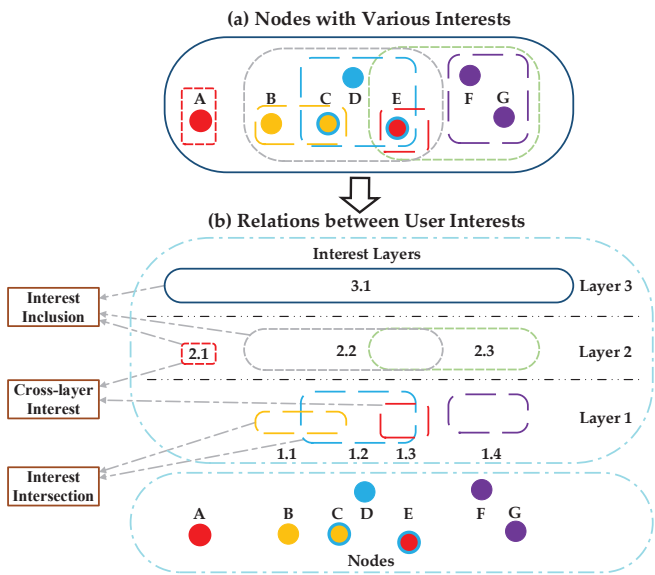}
\caption{Transition of user interests in part (a) to a layered structure in part (b). The figure illustrates the relations between user interests, including interest inclusion, cross-layer interests, and interest intersection.}
\label{relations}
\end{figure}

Fig. \ref{relations} illustrates the transition from nodes (i.e. users) with multiple interests (Fig. \ref{relations}(a)) to a layered structure (Fig. \ref{relations}(b)). In Fig. \ref{relations}(a), there are seven nodes (A, B, C, D, E, F, G) with different interests. Nodes with the same interest have the same color, such as Nodes F and G. If a node (e.g. Node C) is featured with a different color in its border, it has at least two interests. The nodes with similar interests are grouped into dashed or solid rounded rectangles. For example, Nodes B, C, D and E are grouped in the gray dashed rounded rectangle while all the seven nodes are in the solid rounded rectangle. In Fig. \ref{relations}(b), there are 3 layers of interests. All the interests in layers 1-3 are numbered for simplicity. The figure helps to introduce the relations between user interests, including inclusion, cross-layer and intersection.

In the rest of this section, we borrow some concepts from the ACM Computing Classification System (CCS) \footnote{http://www.acm.org/about/class/2012.} to explain the various kinds of relations between user interests shown in the left part of Fig. \ref{relations}(b). However, we are aware that ACM CCS is not enough to describe the complexity about user interests and their relations. For example, users are featured with multiple interests and the interests are related to each other. Hence, ACM CCS provides its value here as an instance. The detailed discussion on multiple interests is presented in Section \ref{discussion}.

\subsection{Interest Inclusion}
Interest inclusion is a special kind of relation between user interests in which individuals have different interests with diverse scopes and some interests contain others entirely. The rounded rectangles in layers 1-3 in Fig. \ref{relations}(b) shows the relation of interest inclusion. For example, Interest 3.1 contains all the other interests, and Interest 2.2 contains Interests 1.1-1.3. Additionally, this can be supported by ACM CCS. The keywords in ACM CCS refer to different interests, such that, broader-scope interest like \textit{architectures} covers more specific ones like \textit{serial architectures} and \textit{parallel architectures}, and then the much specific ones (e.g. \textit{multiple instruction}, \textit{multiple data}).

\subsection{Cross-layer Interests}
Cross-layer interests means an interest appears in at least two layers of an interest-tree as shown in Fig. \ref{relations}(b). Interests 2.1 and 1.3 in Layers 2 and 3 are the same interest but appear in two layers. This is also reflected by ACM CCS in the case that \textit{sensor networks} is listed in both \textit{Computer systems organization/Embedded and cyber-physical systems} and \textit{Networks/Network types/Cyber-physical networks} which are different layers.

\subsection{Interest Intersection}
Interest intersection can be understood as overlap. For instance, a node's interests cover \textit{mobile networks} and \textit{ad hoc networks}. According to CCS Concept, these areas are relevant to each other, meaning they share some features in common. Consequently, this is a situation of interest intersection, as what is shown by Interests 1.1 and 1.2 in Layer 3 of Fig. \ref{relations}(b).

\section{Design of Int-Tree}

\subsection{Overview}
\begin{figure}[!ht]
\centering
\includegraphics [width=3in]{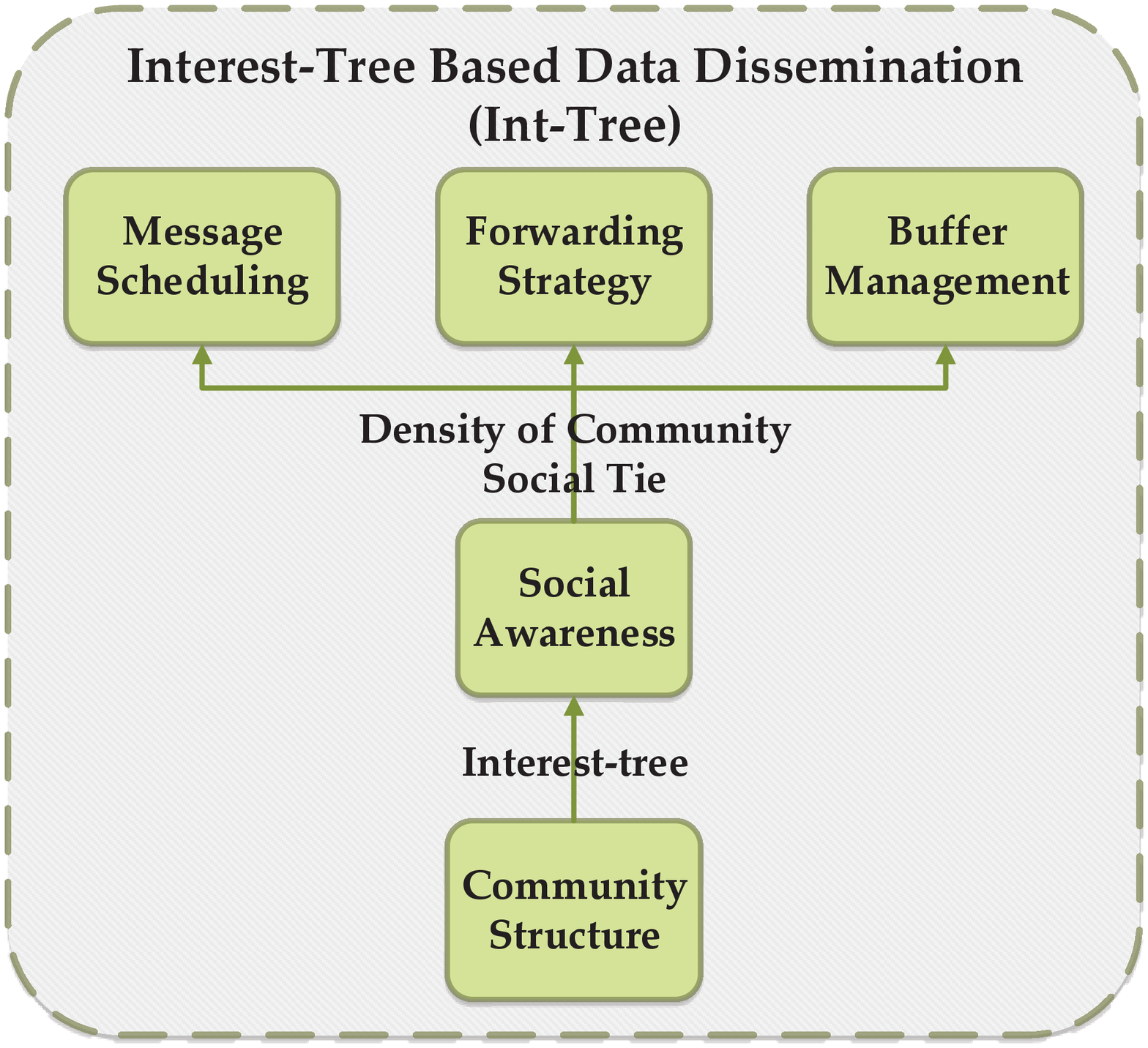}
\caption{Components of Int-Tree.}
\label{components}
\end{figure}

Int-Tree makes use of user interests to construct social communities. Nodes constitute a community when they share the same interests. The components of Int-Tree are depicted in Fig. \ref{components}. From this figure, Int-Tree includes \textit{community structure, social awareness, forwarding strategy, message scheduling, and buffer management} modules. The community structure is for constructing interest-tree. The structure contains interest information of communities which is of crucial importance in presenting the density of communities, and social tie information of social awareness module. Social awareness is responsible for perceiving information of density and social tie. The available social information of nodes make it possible for Int-Tree to choose the appropriate relay nodes and handle efficient message scheduling as well as buffering mechanism. The forwarding strategy takes different strategies to select the next forwarders inside and between the communities. We note that Int-Tree directly applies message scheduling and buffer management modules from BEEINFO \cite{Xia:BEEINFO}.

During the forwarding process, some social information of network such as the community, density of the community and social tie are maintained by nodes. The available feature makes our forwarding scheme flexible with the absence of infrastructure in SAN which is adaptable to more dynamic environment. In the following subsections, detail description for each feature is presented.

\subsection{Social Awareness in Int-Tree}
\label{soctie}
\subsubsection{Community Structure}
\label{structure}
In Int-Tree, we consider user interests as their social features to form the community structure. In other words, nodes with the same interest form a community. Because of node mobility, nodes in the same geographical area may belong to different communities. Hence, Int-Tree detects communities based on interests, instead of space which is more robust when the parameters of network environment change quickly.

\begin{figure}[!ht]
\centering
\includegraphics [width=3.5in]{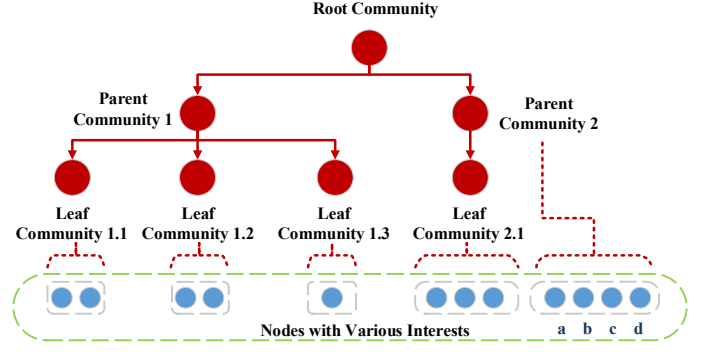}
\caption{An example of the tree-based community structure (interest-tree).}
\label{tree}
\end{figure}

In Section 4, we analyzed the different kinds of relationships between user interests. All these relations contribute to construct a community structure. Intrigued by ACM CCS and researches in \cite{Zhu:Ripple} and \cite{Ruiz:Information} we adopt tree to build interest-tree as the community structure. Because each node has at least one interest and nodes with the same interest form a community, a node can belong to different communities. We regard this as multi-interest situation. It will raise more questions and make our problem more complicated (See Section \ref{discussion} for detailed discussions). To simplify the problem, we adopt a concept, \emph{major interest}, to refer to a users major research interest. This reflects the reality since each academic scholar devotes most of his/her time and energy on a major research area. Therefore, if there is no specified statement, the problem remains a single interest problem. We treat interest and community as one concept in this paper for simplicity. Thus in the interest-tree, each node corresponds to a community, as shown in Fig. \ref{tree}. Generally, there are several layers in an interest-tree. A root community with the largest scale is on the top, then parent communities with medium scales in the middle layers and leaf communities with smallest scales. A node can belong to either a parent community or a leaf community, as a result of his/her preference on describing research interests. Considering the limit of space, only one layer of parent communities is chosen for clearance.

In an interest-tree, each node $i$ belongs to a community $C_i$. For two nodes $i$ and $j$, there is a common parent community $C_{i\&j}$. We define $ComSeq(C_i\rightarrow C_j)$ as a community sequence from community $C_i$ to $C_{i\&j}$ ($ComSeq(C_i\rightarrow C_{i\&j})$) at first and then from $C_{i\&j}$ to community $C_j$ ($ComSeq(C_{i\&j}\rightarrow C_j)$), along the interest-tree. Take Fig. \ref{tree} as an example, the community sequence between Leaf Community 1.1 and Leaf Community 2.1, $ComSeq(1.1\rightarrow2.1)$ is $\{1.1, 1, Root~Community, 2, 2.1\}$ (some words omitted due to space restrain). Similarly, $ComSeq(1.1\rightarrow1.2)=\{1.1, 1, 1.2\}$.  Compared to plenty of message information, interest information is small enough and costs little storage space, so the constrained resources are saved.

\begin{table}[!ht]
\centering
\caption{Definition of symbols}
\label{symbol}
\begin{tabular}{l l}                                                            \hline
Symbol                          & Description                               \\  \hline
$i, j$                          & Nodes $i$, $j$                            \\ 
$C$                             & Community                                 \\ 
$C_i$                           & Community of node $i$                     \\ 
$C_{i\&j}$                      & \tabincell{l}{Common parent community of \\ nodes $i$ and $j$}            \\ 
$ComSeq(C_i\rightarrow C_j)$    & Community sequence                        \\ 
$M$                             & Message                                   \\ 
$SN$                            & Source node                               \\ 
$IN$                            & Intermediate node                         \\ 
$DN$                            & Destination node                          \\ 
$I_{sn}$                        & Interest of source node                   \\ 
$I_{in}$                        & Interest of intermediate node             \\ 
$I_{dn}$                        & Interest of destination node              \\  \hline
\end{tabular}
\end{table}

We would like to make some points clear about interest-tree after formal construction. At first, the information of user interest is obtained before the construction of interest-tree, and the dissemination process. Each node locally stores the interest-tree, the interests of its own and the contacted nodes' interests. Besides, each node also locally conducts the calculation about social tie and density as they can meet others with new interests.

Secondly, we intend to integrate the scheme into a conference managing platform, in which the tree is constructed in a centralized way when the server collects the information uploaded by users (attendees). Users do not need to acquire the tree structure in real time. Instead, it is the server's responsibility to push the latest version to clients regularly. This is determined by the conference scenario, in which we just have to guarantee the structure obtained by all the terminals are the same at a certain time point. Hence, there is no need to notify users neither the change of the interest-tree nor the dissemination process, as long as users have some memory space to store a copy of the tree and pay for the traffic consumption of information uploading, interest-tree acquisition and subsequent updates.

Following these points, we prefer to apply a fixed rule to construct the tree and focus more on the advantages of such tree structure on data dissemination. We summarize notations/symbols definition in Table \ref{symbol} and state the necessary assumptions used through out this paper:

\begin{itemize}
\item Each node is responsible for maintaining the same tree structure of interest;
\item Nodes follow their regular mobility patterns;
\item A forwarding message has the same interest tag with the destination node.
\end{itemize}

\subsubsection{Density of Community}
\label{Density}
We define density of a community as the number of passing nodes with different interests (or communities). That is, the bigger the density, the more nodes can be met. Providing that individuals have regular mobility patterns, the density information can be used to select forwarders. Specifically, when the nodes move around, they calculate the density of different communities by counting the pass-by nodes. To the same community, different nodes may have different densities about it. For a certain node, the larger density of a community it has, the more nodes it encounters. In inter-community process, density can help in selecting forwarders to deliver the messages.

In social networks, the degree of sociability is widely utilized to predict the number of encounter nodes for a specific node \cite{Vendramin:GrAnt}. We redefine the degree as the density of a community in equation (1):

\begin{equation}
Dst_{i, C_{m}}(t)= n
\end{equation}
where $Dst_{i, C_{m}}(t)$ is the number of nodes ($n$) belonging to $C_{m}$ node $i$ has contacted over a time period $t$. When $t$ is multiple of time window $T$, we apply the exponentially weighted moving average \cite{Vendramin:GrAnt} to calculate the community density for the future time slot $\Delta t$ as shown in equation (2):

\begin{eqnarray}
Dst_{i, C_{m}}(t+\Delta t) &=& (1-\alpha) \times Dst_{i, C_{m}}(t-\Delta t) \times \gamma ^{k}\nonumber\\
                    & &+ \alpha\times Dst_{i, C_{m}}(t)
\end{eqnarray}
where $\alpha$ is community density prediction factor. The evaporation factor $\gamma$, and the time interval towards the last update $k$ are used in last update in order to weaken the influence of the old density $Dst_{i}(t-\Delta t)$.

\subsubsection{Social Tie}
The social tie indicates the strength of social relationships between two nodes. Nodes with more social similarities or contact duration times have strong social ties with one another \cite{Ioannidis:Strength}. If there is a strong social tie between node A and B, it means that the contact probability between them is high. In this situation, when B is the destination of a message, A is an optimal candidate to relay the message to B. In Int-Tree, we use the social tie feature to choose the best forwarder nodes during intra-community forwarding process.

\label{4_1_2_Intra}
When a node is transmitting a message, the social tie between nodes determines the efficiency. The reason is that social tie is constructed during the contact process, and it indicates the direct contact probability of two nodes which can be used to find the destination node directly. We use $SoTie_{i,j}(t)$ to measure the social tie of nodes $i$ and $j$ during a time period $t$,
\begin{equation}
SoTie_{i,j}(t)=CT_{i,j}
\end{equation}
where $CT_{i,j}$ is the contact times of $i$ and $j$ over time period $t$.

Similar to the calculation of density of community, when $t$ is multiple of a time window $T$, we conduct an evaporation process when combining the past and present values of social tie in order to predict the future social tie, as equation (4):

\begin{eqnarray}
SoTie_{i,j}(t+\Delta t)&=&(1-\beta ) \times SoTie_{i,j}(t-\Delta t) \times \gamma ^{k}\nonumber\\
                       & &+ \beta \times SoTie_{i,j}(t)
\end{eqnarray}
where $\beta$ is the prediction factor of social tie.

When $i$ and $j$ contact each other, they update the density or social tie parameters of each other. The update has no relation to messages, and only involves $i$ and $j$. Algorithm 1 presents the pseudocode of update procedures of community density and social tie. We note that this procedure is described from $i$'s perspective.

\begin{algorithm}
  \caption{Pseudocode of updating community density and social tie}
  \begin{algorithmic}[1]
  \ForAll {$j$ connected to $i$}
    \State //In time period $t$:
    \If {$I_{i}==I_{j}$}
        \State //Update contact information
        \If {$i$ has no social tie record of $j$}
            \State //Initiate $i$'s social tie to $j$;
            \State $CT_{i,j} \leftarrow 0$;
        \EndIf
        \State //Count the contact time of $j$;
        \State  $CT_{i,j} \leftarrow CT_{i,j}+1$;
    \EndIf
    \ForAll{$C$ $\in~ComSeq(C_{j}\rightarrow C_{root})$}
        \State //Update inter-community contact information
        \If {$i$ has no density record to $C$}
            \State //Initiate $i$'s density to $C$;
            \State $Dst_{i, C} \leftarrow 0$;
        \EndIf
        \State //Count the number of $SN$'s contacted nodes in $C$;
        \State $Dst_{i, C} \leftarrow Dst_{i, C}+1$;
    \EndFor
    \State //When $t$ is multiple of $T$
    \State //Compute density and social tie information:
    \State Compute $Dst(t+\Delta t)$ using Equation (2)
    \State Compute $SoTie(t+\Delta t)$ using Equation (4)
  \EndFor
  \end{algorithmic}
\end{algorithm}

\subsection{Forwarding Strategy}
The forwarding module is the core of Int-Tree. It combines interest-tree, density of community, and social tie to choose the best forwarder nodes. The destination information, such as ID and interest, is included in the message which make it easy to obtain the corresponding information of the destination node. According to the interests of $DN$, $SN$ and $IN$, Int-Tree classifies the environment context into inter-community and intra-community parts, and takes some criteria in different conditions to measure them.
\begin{itemize}
\item $I_{SN}$ == $I_{DN}$ and $I_{IN}$ == $I_{DN}$. Under this condition, $DN$, $SN$, and $IN$ share the same interest, so it is intra-community process. Thus, social tie is utilized to decide the better forwarder. The node with a higher social tie to $DN$ will be selected as the forwarder among $SN$, and $IN$. Otherwise $SN$ will stop the process.
\item $I_{SN}$ == $I_{DN}$ and $I_{IN}$ != $I_{DN}$. $DN$ and $SN$ still share the same interest, while $IN$ has a different one. Thus, $DN$ and $SN$ are in the same community with $IN$ as an outsider. This suggests we need a node in the same community with $DN$ to perform intra-community forwarding, so $IN$ is not suitable.
\item $I_{SN}$ != $I_{DN}$ and $I_{IN}$ == $I_{DN}$. $IN$ and $DN$ have the same interest, while $SN$ is distinct. Thus $IN$ and $DN$ are in the same community, while $SN$ is not. Therefore, the inter-community forwarding suits the situation better. The forwarding strategy will select $IN$ as a forwarder and the message will be forwarded from $SN$ to $IN$.
\item $I_{SN}$ != $I_{DN}$ and $I_{SN}$ == $I_{IN}$. In this situation, $SN$ and $IN$ belong to the same community but not the destination community. It is inter-community forwarding. Int-Tree will utilize the density information to make decision. To be exact, $SN$ and $IN$ compare their densities to all the $C$s belonging to the community sequence from $C_{DN}$ to $C_{SN\&DN}$ (i.e. $ComSeq(C_{DN}\rightarrow C_{SN\&DN})$). If $IN$ has the larger density to a certain $C$ in the sequence, $SN$ will choose $IN$ to forward; Otherwise, $SN$ will keep the message.
\item $I_{SN}$ != $I_{DN}$, $I_{IN}$ != $I_{DN}$ and $I_{SN}$ != $I_{IN}$. $SN$, $IN$ and $DN$ share no common interests at all, so they are in different communities (inter-community). Thus, Int-Tree performs the same procedure as the last condition to choose a forwarder.
\end{itemize}

When the destination node receives a message, it broadcasts a response message to all the nodes that have the message in order to help them to drop the message. Algorithm 2 presents the pseudocode of the forwarding strategy. During the forwarding process, the concept of $SN$ and $IN$ are not related to specific nodes. That is, the source node, say $IN$, can be a relay node, say $SN$, in the future.

\begin{algorithm}[t]
  \caption{Pseudocode of forwarding strategy}
  \begin{algorithmic}[1]
    \State Given a message $M$ in the buffer of a node;
    \ForAll {$IN$ contacting $SN$}
        \If {$IN$ is $DN$}
            \State $SN$ delivers $M$ to $IN$;
        \Else
            \If {$I_{IN}==I_{DN}$}
                \If {$I_{SN}==I_{DN}$}
                    \If {$SoTie(SN,DN)<SoTie(IN,DN)$}
                        \State $SN$ delivers $M$ to $IN$;
                    \EndIf
                \Else
                    \State $SN$ delivers $M$ to $IN$;
                \EndIf
            \Else
                \If {$I_{SN}\neq I_{DN}$}
                    \ForAll{$C~\in~ComSeq(C_{DN}\rightarrow C_{SN\&DN})$}
                        \If {$Dst(SN, C)<Dst(IN, C)$}
                            \State $SN$ delivers $M$ to $IN$;
                            \BREAK
                        \EndIf
                    \EndFor
                \EndIf
            \EndIf
        \EndIf
    \EndFor
  \end{algorithmic}
\end{algorithm}

\subsection{Message Scheduling and Buffer Management}

Message scheduling decides the order of transmitting the messages between nodes. It ensures that the messages with higher forwarding opportunities have higher priority for forwarding. On the other hand, buffer management module decides which messages should be discarded when the buffer is overloaded. The message scheduling and buffer replacement share similar principles, as they both require excluding or discarding the expired or successfully delivered messages without influencing those messages which are transmitting. We suggest readers who are interested to refer to \cite{Xia:BEEINFO} for more details about these components.

\section{Performance Evaluation}
\begin{table*}
\centering
\caption{Statistics in Regard of the Interest-Tree}
\label{stat}
\begin{tabular}{l l l l l}                                                            \hline
Layer No.   & Involved Interest No.                     & Sum (Interest)    & Involved Node No.                 & Sum (Node)        \\ \hline
1           & 0                                         & 1                 & No nodes involved.                & 0                 \\ 
2           & 1 60                                      & 2                 & \tabincell{l}{2 3 4 5 6 7 9 14 15 16 18 20 25\\ 27 28 29 31 32 36 40 41 42 44 52\\57 67 68 71}  & 28                \\ 
3           & \tabincell{l}{2 3 4 5 6 8 9 10 13 28 30 31 34 37 39\\40 59 78 79}    & 19                & 0 21 43 51 55 61 69 70                    & 8                 \\ 
4           & \tabincell{l}{7 11 12 14 15 16 17 18 19 22 23 26 27\\35 36 38 41 43 45 46 49 52 53 54 55\\56 58 61 62 68 70 71 74 76 77 80}     & 36                & \tabincell{l}{1 10 11 13 17 23 30 33 34 35 37\\39 45 46 54 58 59 63 65 66 72 74}        & 22                \\ 
5           & \tabincell{l}{20 21 24 25 29 32 47 48 51 63 64\\65 66 67 75 81 82}          & 17                & 19 22 26 47 48 49 50 53 64 73 75          & 11                \\ 
6           & 33 42 44 50 69 72 73                      & 7                 & 8 12 24 56 60 62                          & 6                 \\ 
7           & 57                                        & 1                 & 38                                        & 1                 \\  \hline
\end{tabular}
\end{table*}
\subsection{Dataset}
\label{Dataset}
The data set adopted in our work was collected during ACM SIGCOMM 2009 \cite{SIGCOMM2009}, available on the website of CRAWDAD (http://www.crawdad.org/thlab/sigcomm2009/). It includes a list of the participants' original interests. The dataset contains the activity periods of each participant and device, as well as the Bluetooth device discovery logs of each user. Therefore, we can get the record of all the meetings between all the participants for simulation experiments, instead of applying any synthetic movement model.

\subsection{Interest-Tree Construction}
\label{Interesttree}
The collected data set during ACM SIGCOMM 2009 includes the tested attendees' interest information and contact information. To be specific, there are 76 attendees (nodes) and 711 interests. We numbered the nodes from No. 0 to No. 75 and the interests from No. 1 to No. 711. However, we found out that some interests involve the same nodes. For example, both Interests No. 79 and No. 82 involve Nodes No. 8 and No. 62. To simplify the data set process, we merge the interests which involve the same nodes into one interest, deriving 82 interests, numbered from 1 to 82.

Because we have decided to use nodes' \textit{major interest}s to conduct research, it is necessary to find a node's \textit{major interest}. Hereby, we denote node $i$'s \textit{major interest} as the interest with the largest number in $i$'s interest list. In this sense, it is appropriate to regard this as a procedure seeking a node's most distinguished interest. After the process, all the nodes's major interests cover 49 of the 82 interests, meaning that some nodes share the same one. Then, we find parent interest for each interest from Interest No. 82 in descending order. If the involved nodes of Interest No. $m$ contain entirely those of Interest No. $n$ ($m \in \{n-1, n-2, ..., 1\}$), No. $m$ is the parent interest of No. $n$. For Interests No. 1 (involving all the nodes except for Node No. 44) and No. 60 (involving only Node No. 44), we create Interest No. 0 as their parent interest. Finally, an unbalanced 7-layer multi-tree is generated, with 1 root interest, 35 parent interests, and 47 leaf interests. Table \ref{stat} shows the statistics in regard of the tree.

While constructing interest-tree, we merged the interests involving the same nodes, and chose $I_{i}$ manually. These actions do not fully reflect the reality and may induce biased experiment results. However, these are inevitable due to the missing information of users' real interests in the data set. The incompleteness of the data set is probably the result of concerns about information security or privacy issues. Therefore, the experiment result here is compromised regarding the above issues.
\subsection{Simulation Setup}
\label{simulationSetup}
\begin{table}
\centering
\caption{Simulation parameters}
\label{Parameter}
\begin{tabular}{l l l}                                                               \hline
Simulation Parameters       & Values                              & Default Value  \\  \hline
Simulation time             & 21600$\scriptsize{\sim}$129600 s    & 216000 s        \\ 
Number of runs              & 10                                  & -              \\ 
Warm up time                & 1000s                               & -              \\ 
Time window $T$             & 30s                                 & -              \\ 
Network Area                & $80\times80$ m$^{2}$                & -              \\ 
Wait time at destination    & 0$\scriptsize{\sim}$120 s           & Random         \\ 
Event interval              & 25$\scriptsize{\sim}$35 s           & Random         \\ 
Message size                & 500$\scriptsize{\sim}$1024 KB       & Random         \\ 
Message TTL                 & 60$\scriptsize{\sim}$360 min        & 360 min        \\ 
Node buffer size            & 5$\scriptsize{\sim}$30 MB           & 5 MB           \\ 
$\alpha$                    & 0.7                                 & -              \\ 
$\beta$                     & 0.1                                 & -              \\ 
$\gamma$                    & 0.9                                 & -              \\  \hline
\end{tabular}
\end{table}

Our experiments are carried out using the Opportunistic Network Environment (ONE) simulator \cite{Keranen:Simulating}. We chose the Epidemic routing and PROPHET algorithm as the benchmark DTN routing protocols (please see Section 2 for detailed reasons). We run our simulation for the three protocols with different simulation time (21600s$\scriptsize{\sim}$129600s), buffer sizes (5MB$\scriptsize{\sim}$30MB), and message TTL (60min$\scriptsize{\sim}$360min). Default value of parameters are 21600s, 5MB and 360min, respectively. The wireless transmission applies Bluetooth interface with a 10-meter communication range and 2-Mbps transmission speed. The reasons we use Bluetooth interface include: i) it consumes the least energy compared to Wi-Fi or 3G technology; and ii) the transmission range of Bluetooth is enough for our conference scenario where the possible forwarders move in a small area \cite{Chuang:Bluetooth}. Events are generated each 25$\scriptsize{\sim}$35 seconds. The simulation parameters are summarized in Table \ref{Parameter}.

In each experiment, we compare the performance of protocols based on the following metrics:

\textbf{Message Delivery Ratio}: the ratio of successfully delivered messages to the total number of unique messages created (excluding redundant messages) in a given period.

\textbf{Overhead}: the ratio of relayed messages (delivered messages excluded) and delivered messages.

\textbf{Average Latency}: the average time between the time a message is generated and the time it is delivered successfully.

\textbf{Average Hop Count}: the average hop-counts when messages are received successfully.

\subsection{Preliminary Results}

\begin{figure*}[!ht]
\centering
\subfigure[Delivery Ratio]{
\label{fig:5-a}
\includegraphics[width=0.235\textwidth]{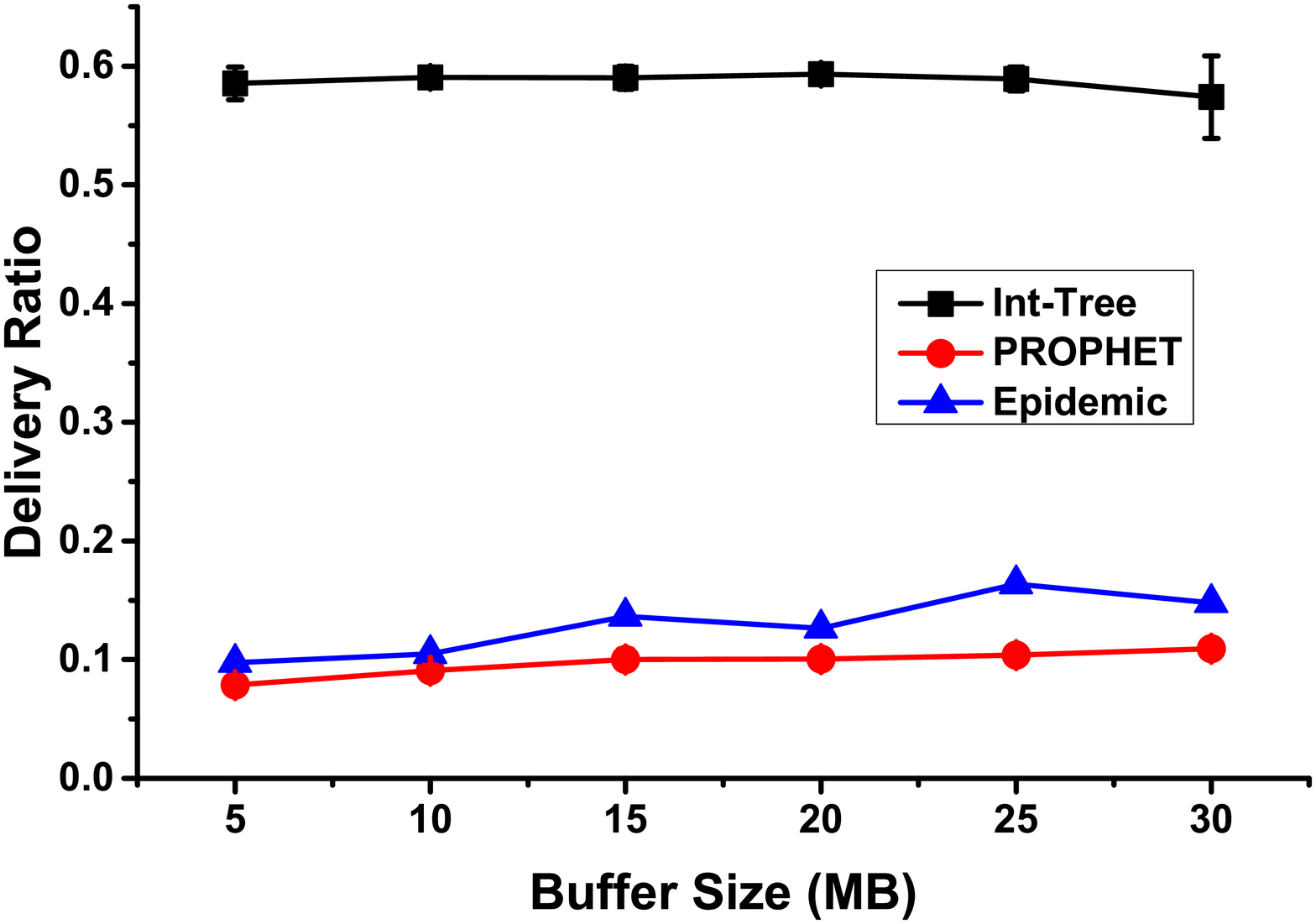}}
\subfigure[Overhead]{
\label{fig:5-b}
\includegraphics[width=0.235\textwidth]{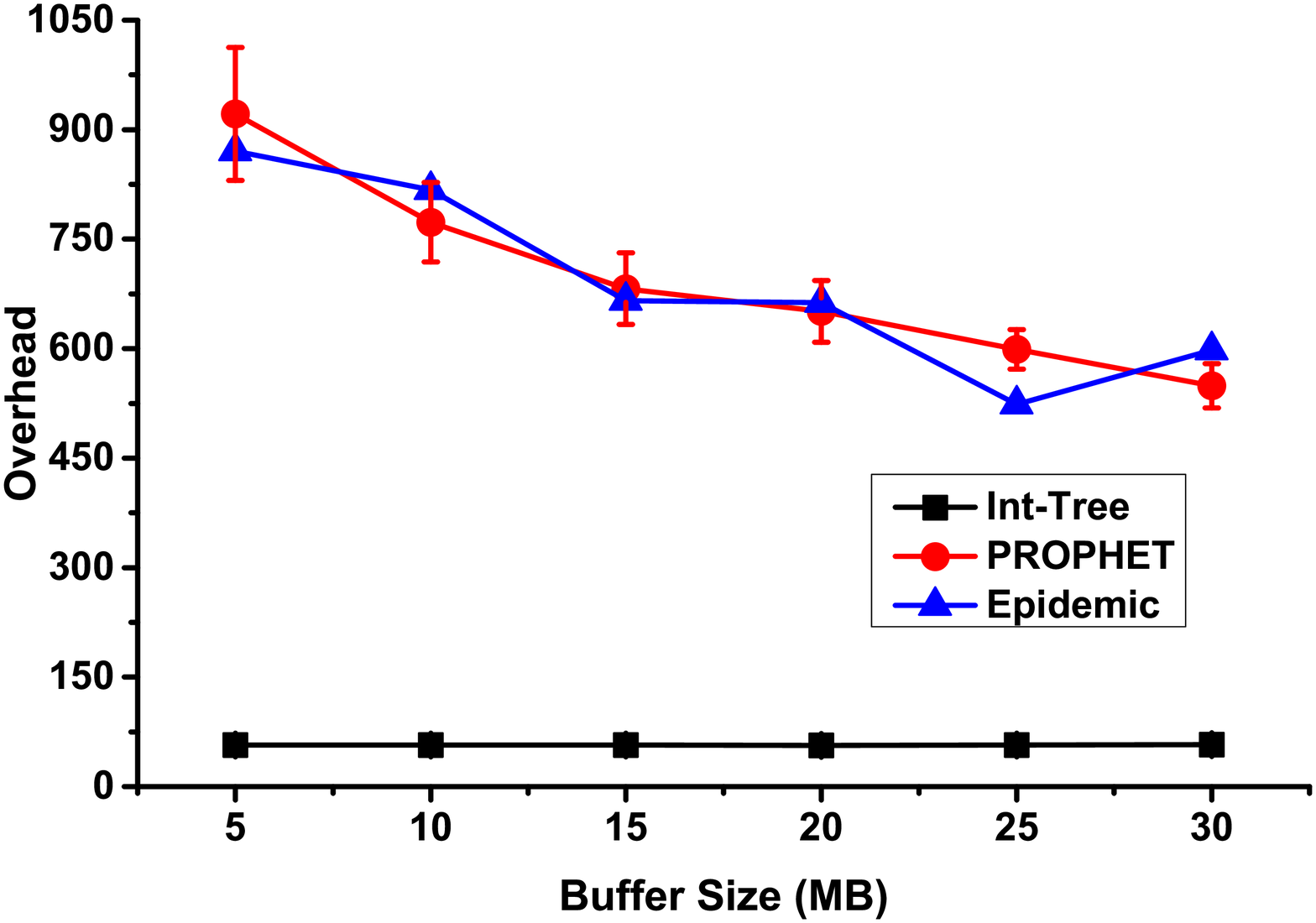}}
\subfigure[Hop-count]{
\label{fig:5-c}
\includegraphics[width=0.235\textwidth]{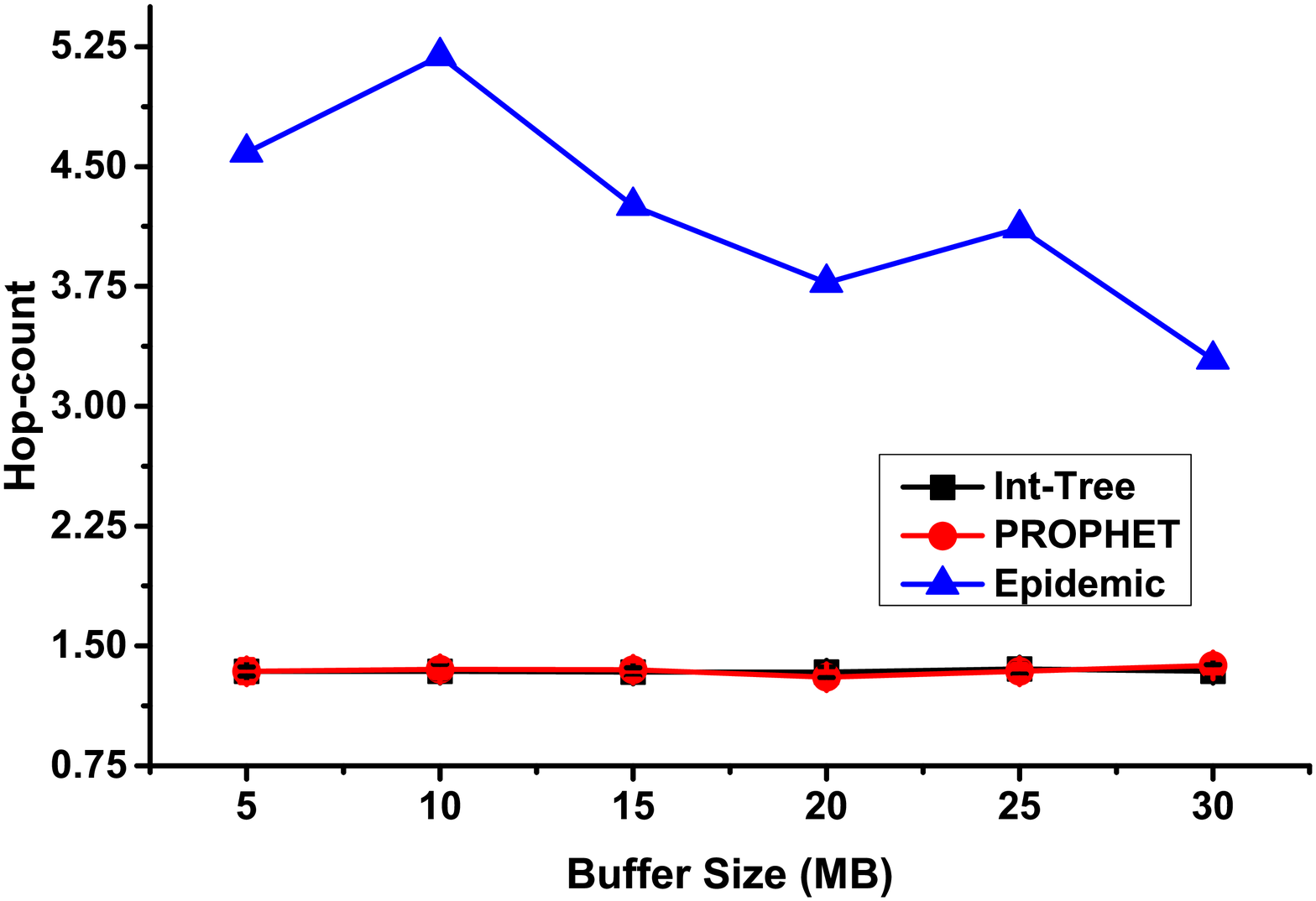}}
\subfigure[Average Latency]{
\label{fig:5-d}
\includegraphics[width=0.235\textwidth]{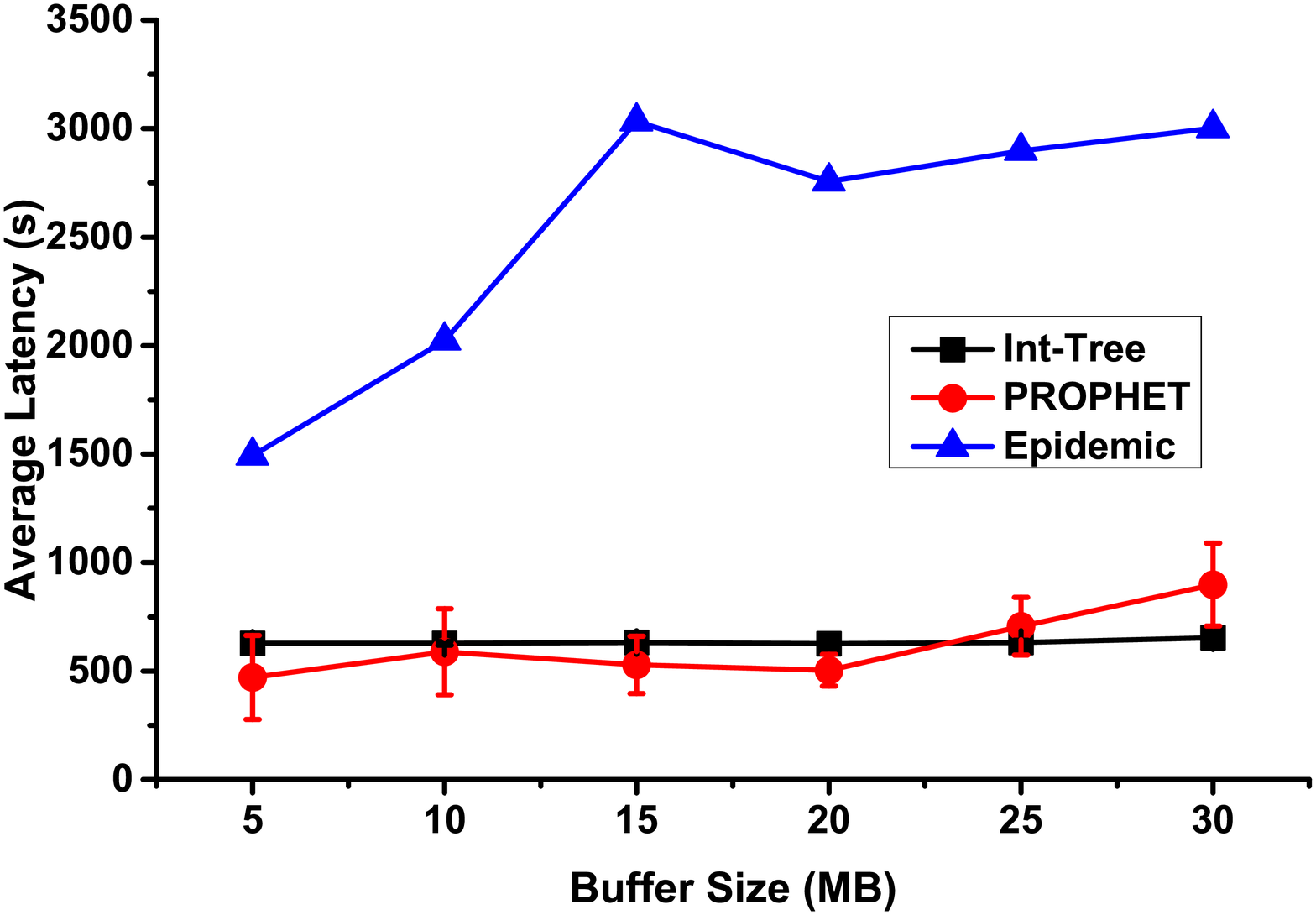}}
\caption{Performance with respect to buffer size.}
\label{buffer}
\end{figure*}

Fig. \ref{buffer} compares the performance of the three algorithms over different buffer sizes. When the size of buffer is risen, both PROPHET and Epidemic show clearly increasing trend for delivery ratio and average latency, and decreasing trend for overhead and hop-count. However, Int-Tree vibrates little for all the parameters. Comparatively, it can be seen that Int-Tree has the best performance in this situation.

Specifically, when the buffer size is set to 20MB (see Fig. \ref{buffer}(a)), Int-Tree forwards 59.32\% messages while the delivery ratios of PROPHET and Epidemic are 10.04\% and 12.63\%, respectively. As shown in Fig. \ref{buffer}(b), the overhead of Int-Tree remains lower than 60, which is much smaller than the other two protocols where their overhead is higher than 500. Comparing the hop count of the protocols in Fig. \ref{buffer}(c), the average value of Int-Tree is about 1.3423, slightly lower than that of PROPHET (1.3449), while the average hop count of Epidemic vibrates dramatically between 3.29 to 5.19. Fig. \ref{buffer}(d) shows a trend that Int-Tree spends more time than PROPHET when buffer size is smaller than 20 MB, but after the 20-MB point, Int-Tree consumes the least time. On the contrary, Epidemic performs the worst for using more than 1500s to disseminate messages. This is because Epidemic generates many copies of message, causing more buffer operation.

For the stable performance of Int-Tree, we believe it is a result of the small amount of information it deals with. In the dataset, both mobile nodes and user interests are in small scale. Therefore, it requires little buffer for both interest-tree and computation. This inspires us to do more experiments with smaller buffer sizes in the future.

\begin{figure*}[!ht]
\centering
\subfigure[Delivery Ratio]{
\label{fig:6-a}
\includegraphics[width=0.235\textwidth]{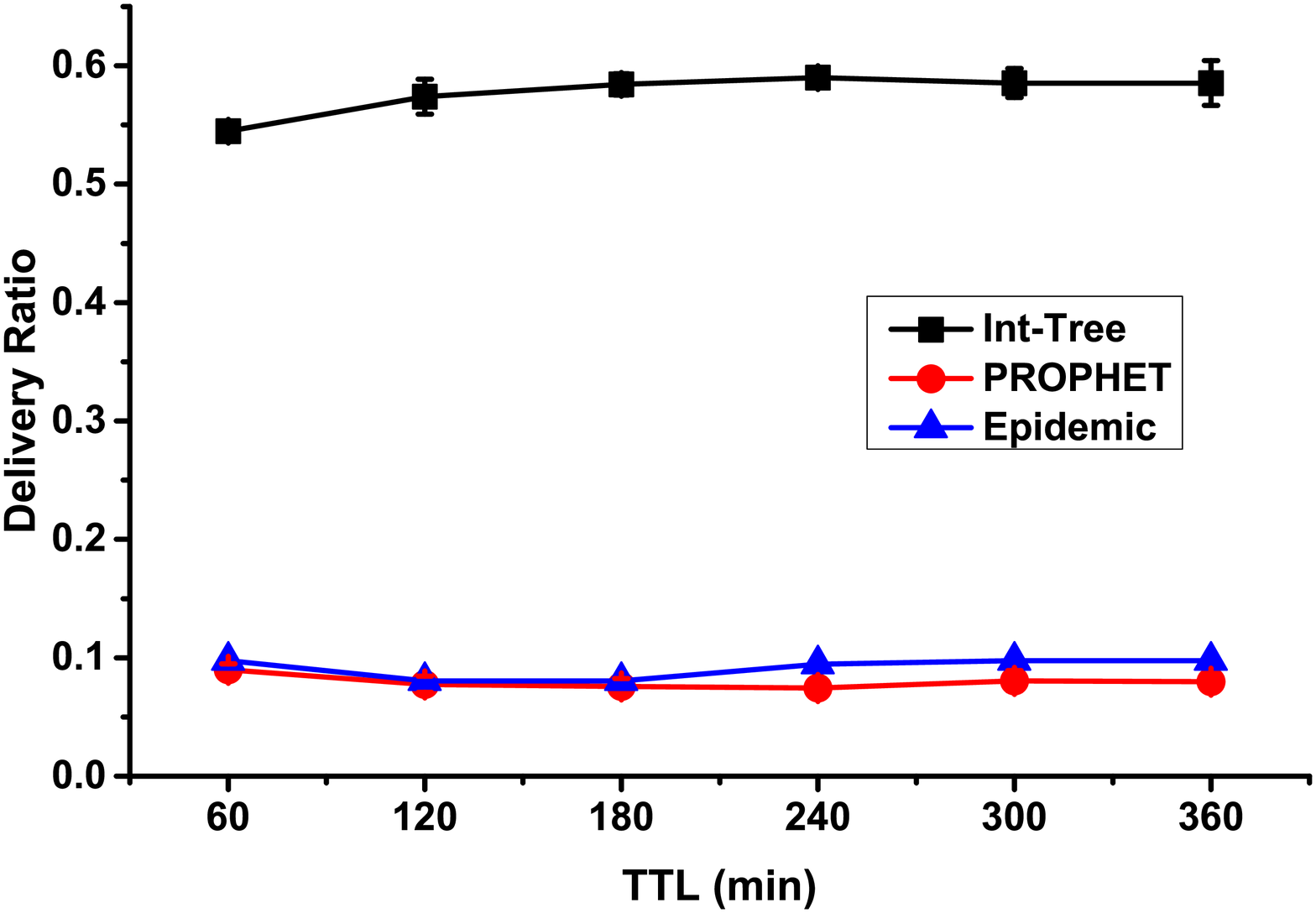}}
\subfigure[Overhead]{
\label{fig:6-b}
\includegraphics[width=0.235\textwidth]{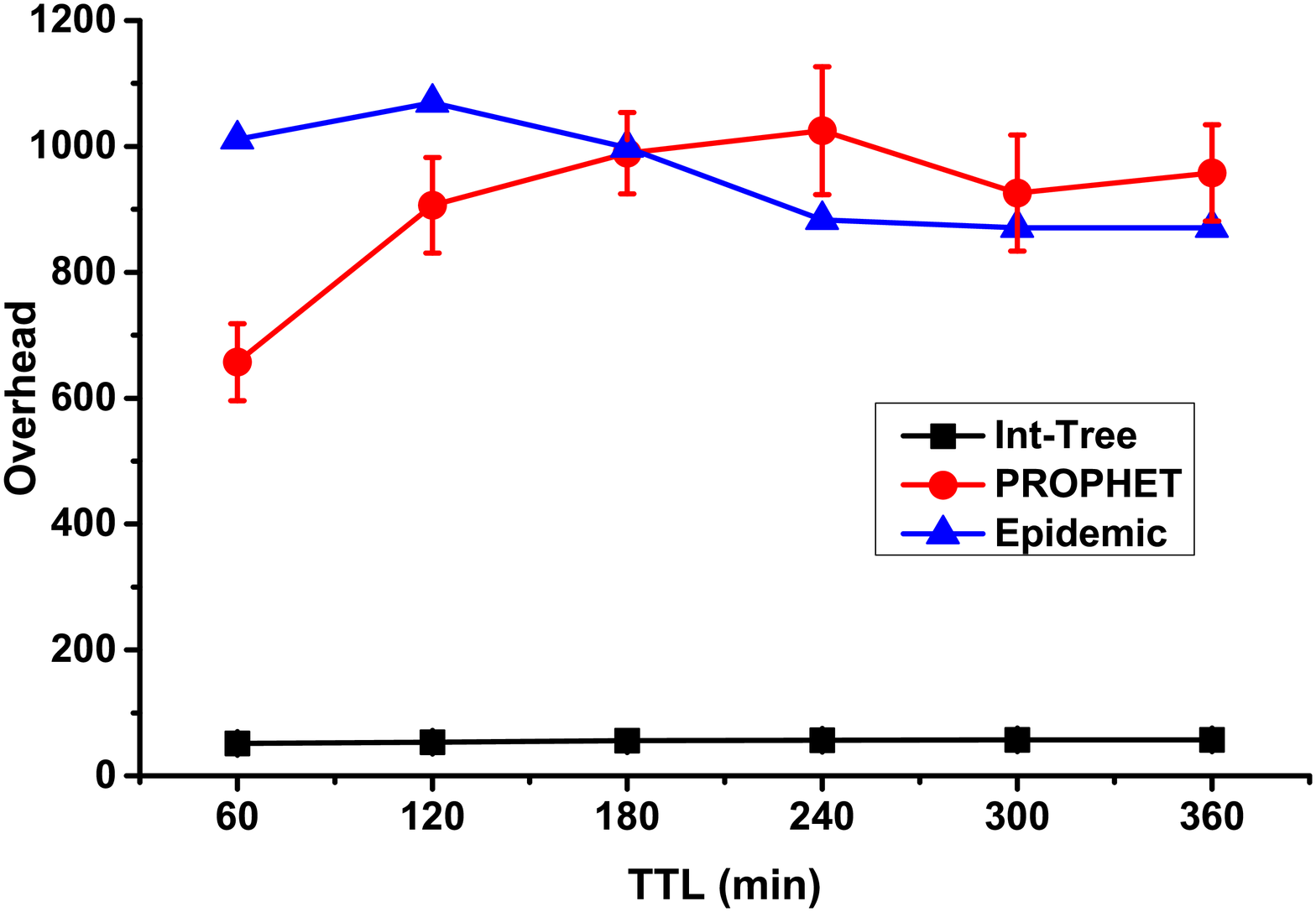}}
\subfigure[Hop-count]{
\label{fig:6-c}
\includegraphics[width=0.235\textwidth]{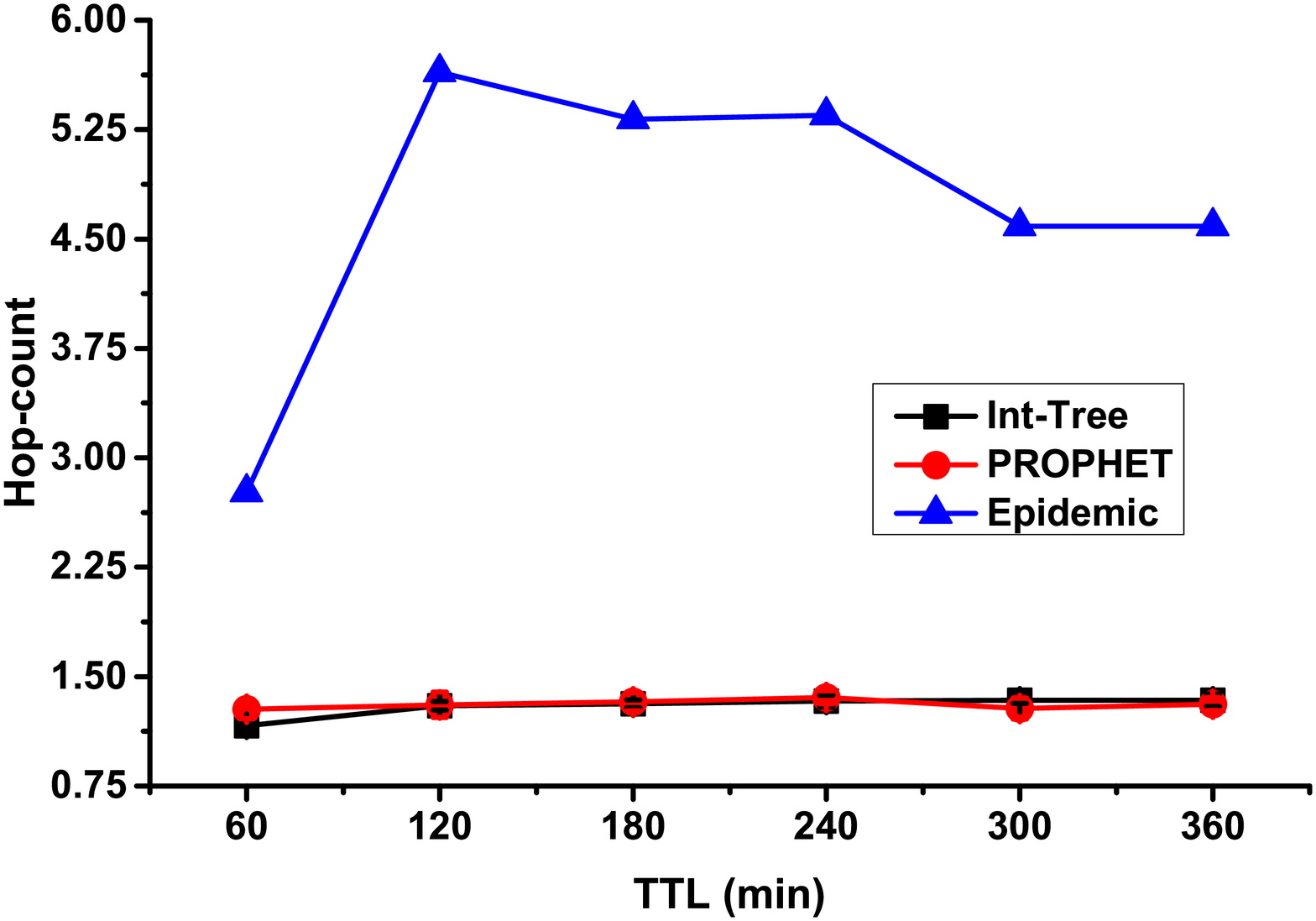}}
\subfigure[Average Latency]{
\label{fig:6-d}
\includegraphics[width=0.235\textwidth]{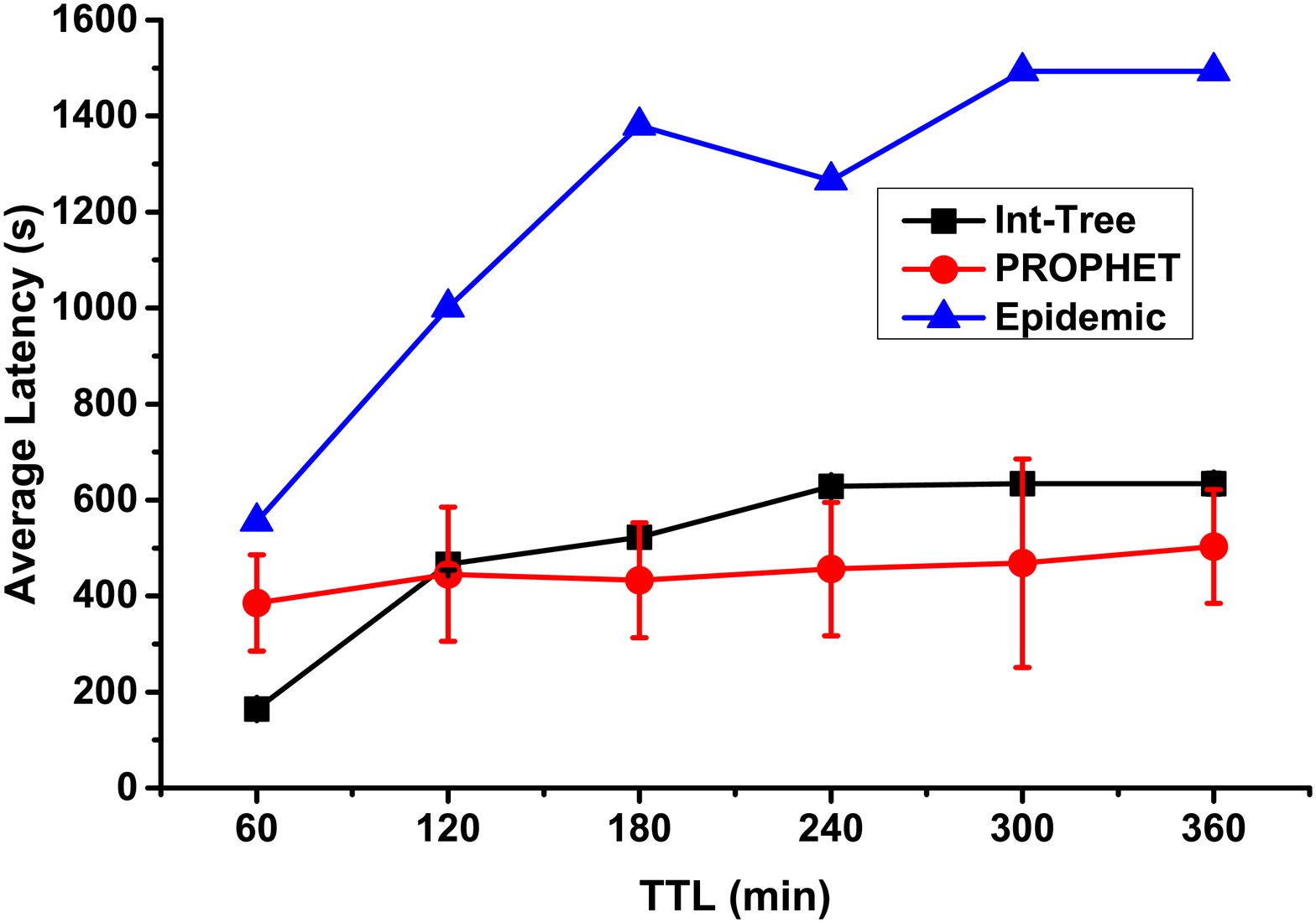}}
\caption{Performance with respect to TTL.}
\label{ttl}
\end{figure*}

Fig. \ref{ttl} compares the performance of the protocols with different values of TTL. According to the figure, it can be seen that performance of almost all the protocols are stable when TTL value of messages changes, except that the Epidemic is more sensitive with respect to hop-count and average latency. In addition, Int-Tree has the best performance with a big advantage in terms of delivery ratio and overhead. For hop-count, Int-Tree remains slightly better than PROPHET. As for average latency (see Fig. \ref{ttl}(d)), Int-Tree spends the least time (164.38s) with 60-min TTL and then rises to higher values than PROPHET. However, the highest value of Int-Tree is 634.31s, 26.1\% higher than PROPHET, 57.5\% lower than Epidemic. \textit{Furthermore, the figure suggests a good trend that Int-Tree is getting stable and the gap between Int-Tree and PROPHET is narrowing.}

\begin{figure*}[!ht]
\centering
\subfigure[Delivery Ratio]{
\label{fig:7-a}
\includegraphics[width=0.235\textwidth]{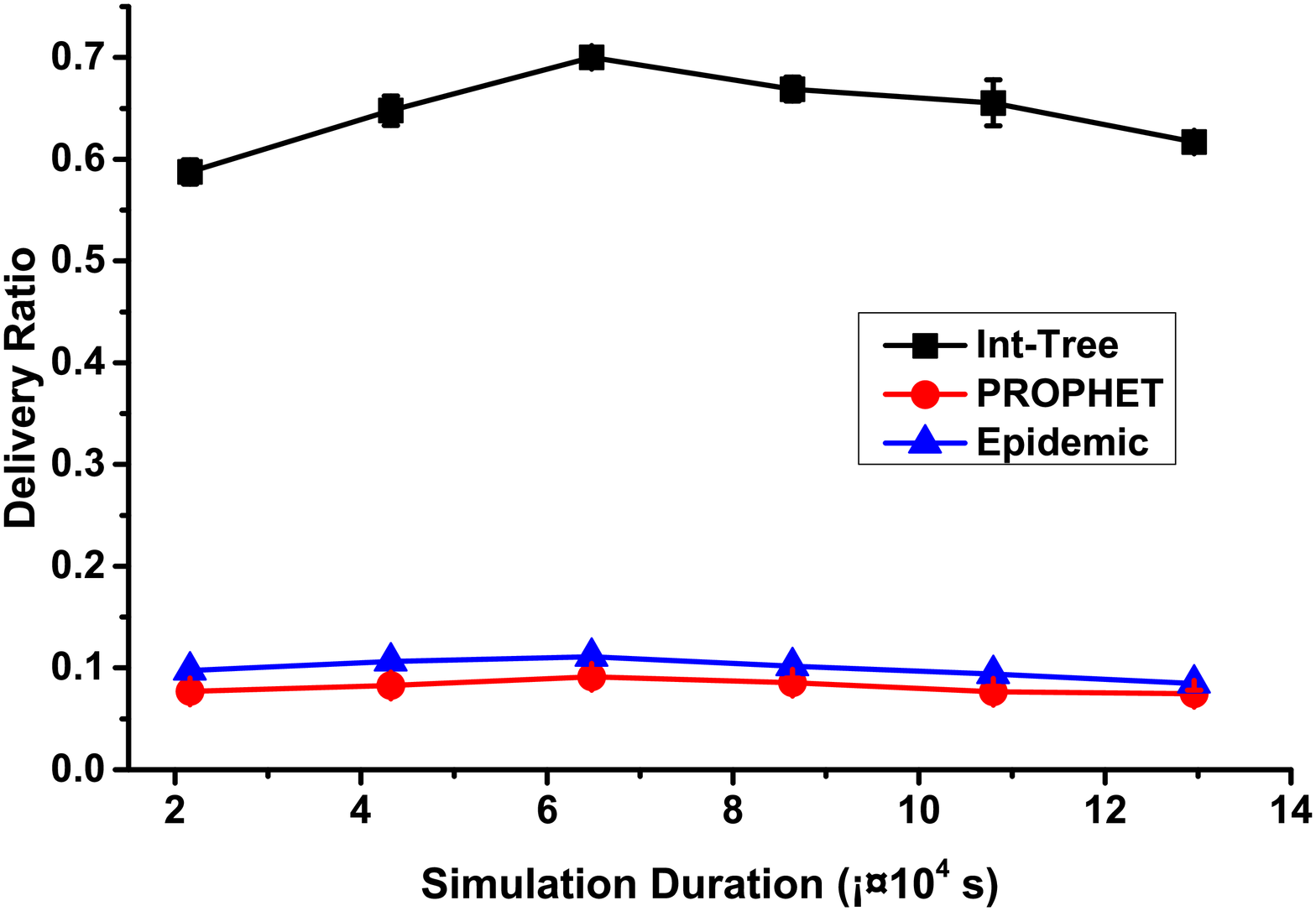}}
\subfigure[Overhead]{
\label{fig:7-b}
\includegraphics[width=0.235\textwidth]{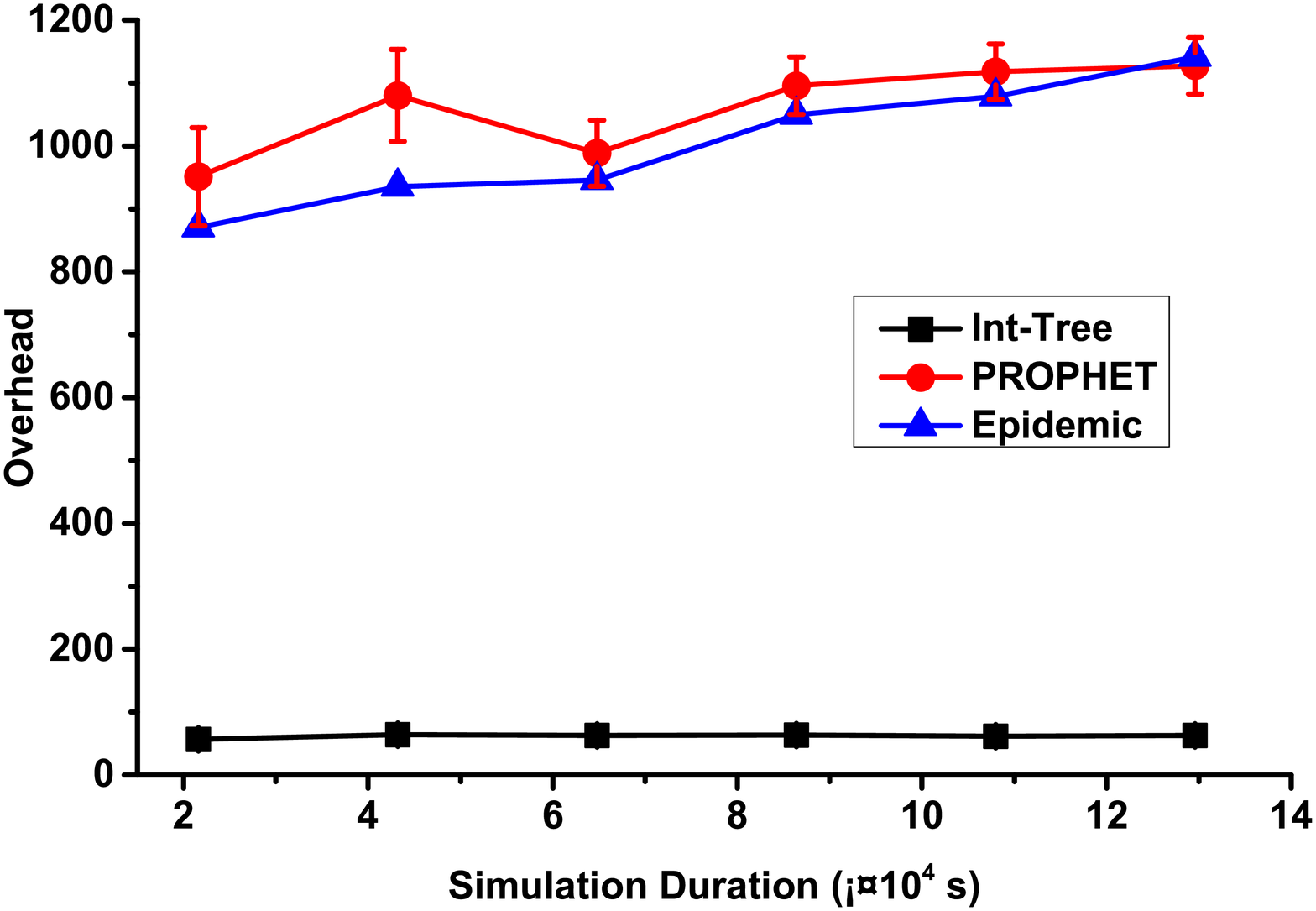}}
\subfigure[Hop-count]{
\label{fig:7-c}
\includegraphics[width=0.235\textwidth]{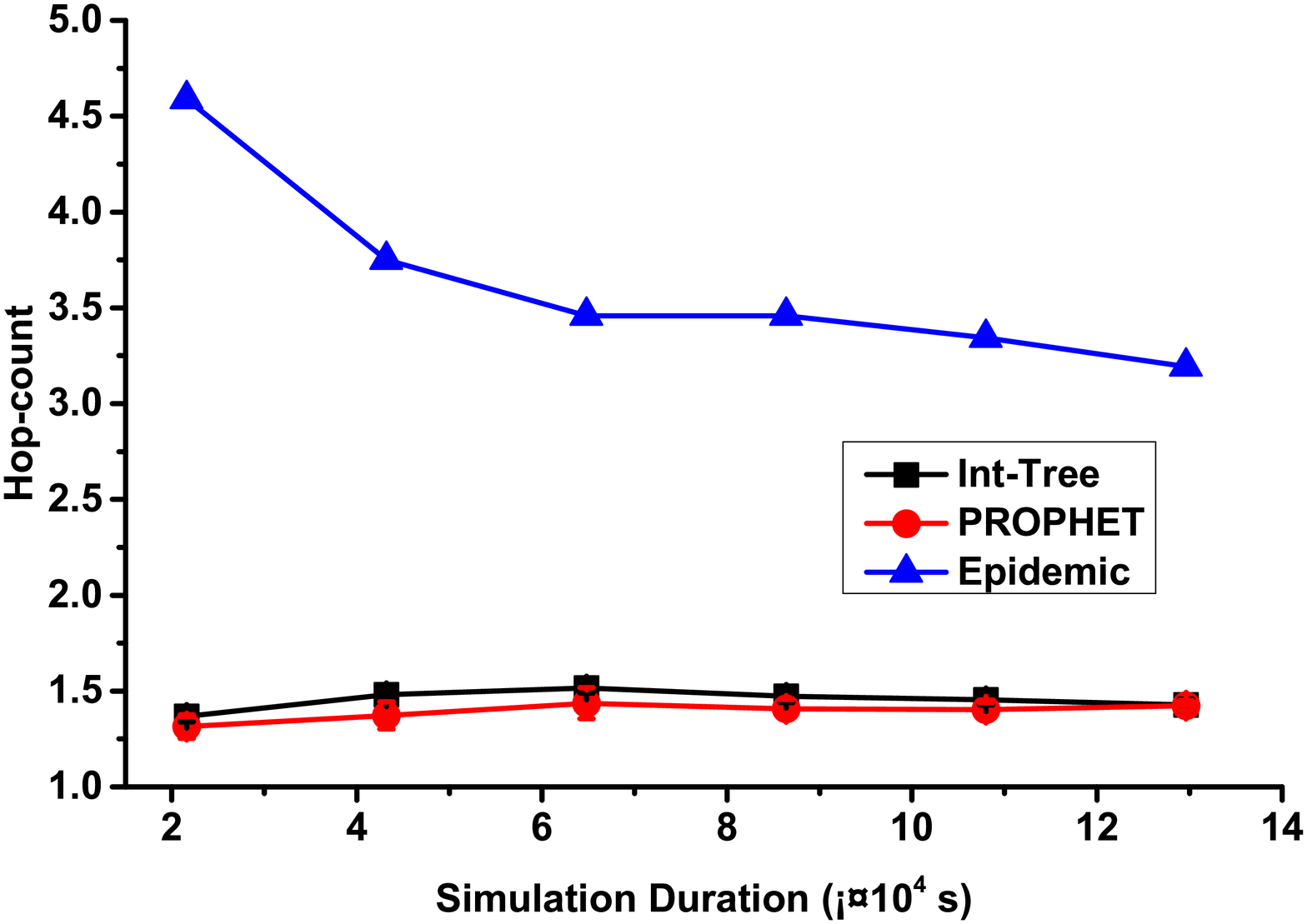}}
\subfigure[Average Latency]{
\label{fig:7-d}
\includegraphics[width=0.235\textwidth]{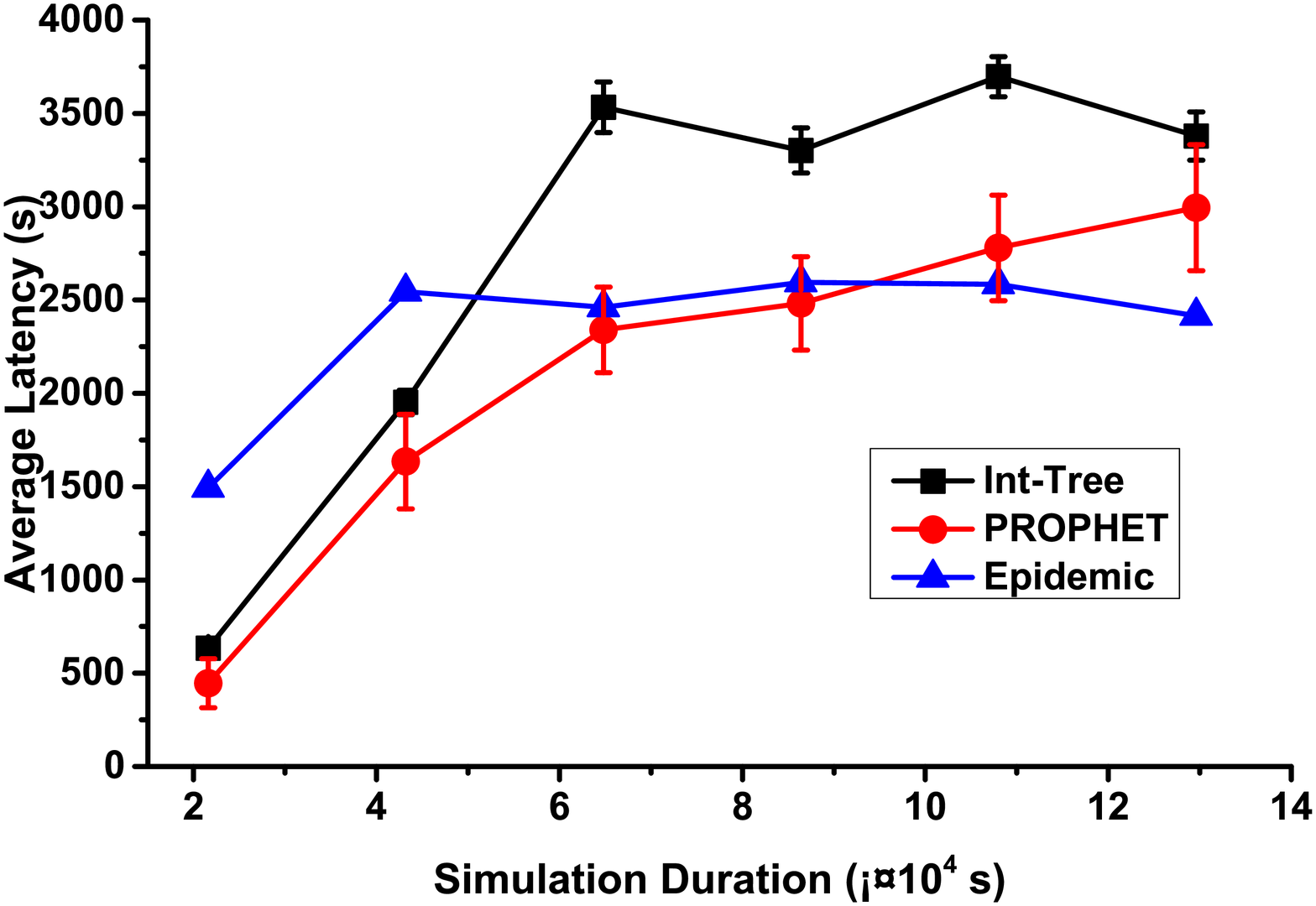}}
\caption{Performance with respect to simulation duration.}
\label{simDuration}
\end{figure*}

Fig. \ref{simDuration} describes the simulation results under different simulation durations. \textit{All the evaluated protocols have experienced similar trend with Fig. \ref{ttl}}, except that: i) both PROPHET and Epidemic keep rising on overhead; ii) it takes less hop-count to disseminate messages for Epidemic when simulation duration rises; and iii) it costs Int-Tree much more time than the others when simulation duration is larger than or equal to 64800s.

The results of our simulations with different values of evaluation parameters show that Int-Tree disseminates data with the highest delivery ratio, the lowest overhead and the minimum hop count. This attributes to that Int-Tree adopts interest-tree, social tie and community density to help make decision on choosing forwarders. Int-Tree deduces unnecessary message copies and transmissions in network which results in low overhead and small hop counts. It also chooses a best forwarder node effectively which increases the delivery ratio. As for the relatively larger latency values, we should notice that it is caused by consuming time to find a proper forwarder. However, we believe this deficiency is tolerable considering that the hope count in our algorithm is less than 1.5 in average.

\subsection{Further Experiments}

We conduct two groups of simulations on exploring how  $\alpha$, $\beta$ and $\gamma$ influence the performance of Int-Tree. $\alpha$ and $\beta$ are both prediction factors, but for different information. Therefore, we design 25 groups of $(\alpha, \beta)$ pairs to examine their influence. The bigger they are, the more important the present information is. $\gamma$ reduces the influence of the history record. The smaller it is, the weaker the history information is. We run the simulations 10 times and other parameters are set default except for simulation duration (21600s). The evaluation criteria remain the same.

\begin{figure}[!t]
\centering
\includegraphics [width=3in]{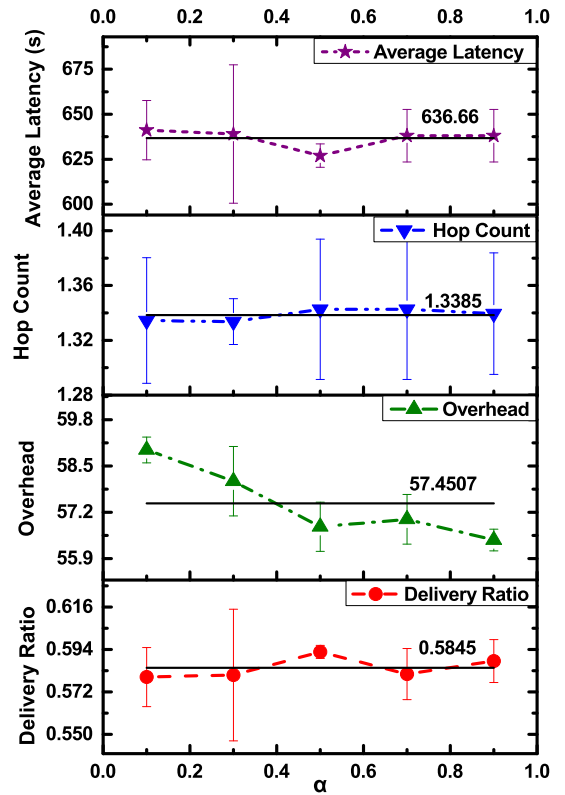}
\caption{Performance with respect to $\alpha$.}
\label{alphabeta}
\end{figure}

\begin{figure}[!t]
\centering
\includegraphics [width=3.55in]{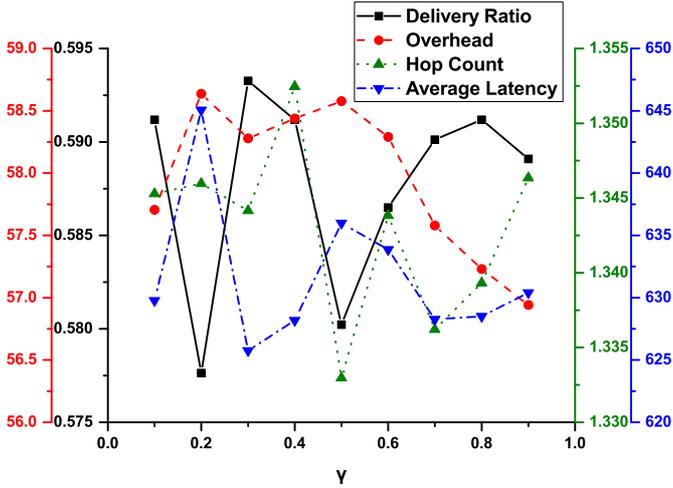}
\caption{Performance with respect to $\gamma$.}
\label{gamma}
\end{figure}

Initially, we expected that $\alpha$ and $\beta$ should have similar impact on Int-Tree. But the result proved a different picture. $\alpha$ does affect the performance of Int-Tree, but no matter how $\beta$ changes, Int-Tree performs the same. \textit{This might be the result of the scarcity of dataset. Nodes belonging to he same community are in small number, making the social tie values between nodes small and constant during the experiment.} Fig. \ref{alphabeta} illustrates the result regarding changing $\alpha$. The black straight lines represent the corresponding average values. From the figure we can see that the values for the metrics are stable and change in small ranges. This suggests that $\alpha$ will not affect Int-Tree significantly. Moreover, we find that $\alpha=0.9$ is the best for Int-Tree because it achieves the second highest delivery ratio, the lowest overhead, the third least hop-count, and the second smallest latency. In other words, a large proportion of present information will help Int-Tree to perform.

Fig. \ref{gamma} shows the performance of Int-Tree over $\gamma$. With $\gamma$ raising, all the evaluated parameters change dramatically, although the ranges are actually very small. Nevertheless, we can still acquire some knowledge from the figure. Generally, delivery ratio experiences the opposite trend from overhead and average latency. This is reasonable because we expect high delivery ratio but low overhead and latency. Moreover, overhead and average latency share the feature of being stable or converging. Hop count shows totally different status with unstable values. Therefore, we need a trade-off to find the best value of $\gamma$ for Int-Tree and this leads us to 0.3. With this value, Int-Tree achieves the highest delivery ratio, the lowest latency, and average performance on overhead and hop-count. In the meantime, it supports the conclusion reached from Fig. \ref{alphabeta} that we need to weaken the influence of the history record.

\section{Discussions on User Interests}
\label{discussion}
As a matter of fact, we considered multi-interest as a relationship between user interests at the beginning of our research. However, we found that it is more of a property of mobile users than of a relationship between interests. This property will make the problem more complicated and the scheme needs improvement in many aspects. Nonetheless, we would like to present some potential issues in regarding of multiple interests and put forward some solutions.
\begin{itemize}
\item \textit{Will the interest-tree change?} Interest-tree is the base to study the relationships between user interests. Introducing multi-interest will not affect the construction rule, the depth and the value of interest-tree. It will only expand the width. The reason behind this is that the emerged interests include the same nodes. To be specific, interest-tree remains 7 layers after emergence, but the width decreases after emergence, proved by the declined number of interests involved in Layers 2-7 shown in Table \ref{width}. Additionally, we ran the experiment with full interest information (i.e, without emergence), and found that the result do not change much, compared to the that with emerged interests. We do not provide the result because Int-Tree needs modification to fully apply to multi-interest situation, and it needs further exploitation on the change of its performance.
\item \textit{How to collect users' interest?} As stated in Section \ref{Interesttree}, the ACM SIGCOMM 2009 dataset did not provide the users' real interest information. While in real life, we should pay attention to the questions concerning interest information collection. For example, will the users provide their interest information, how to provide, or even where exactly to put a specific interest in the tree? In fact, we plan to plant the scheme into a conference platform, in which, there is a module for collecting user information at attendee's will. Users can choose their interests from some drop-down menus. The developers will select an construction algorithm to build the tree, ensuring that the structure is stable, flexible and accessible. The algorithm should be able to distinguish the interests in context, to categorize them and to construct the tree. Moreover, We need to expand ACM CCS structure to a broader scope than computer science. Therefore, techniques in context recognition, or machine learning, or even some inter-discipline ones are required to address this problem.
\item \textit{Will Int-Tree change?} Int-Tree should be changed in both updating social information and choosing forwarders. A preliminary solution is to allow nodes $i$ to check its interest list and update the information for each of its interest. When choosing forwarders, $SN$ needs to check both its shared interests with $IN$ and the distinct interests from $IN$. Furthermore, a special case draws our attention. Node $A$ is interested in a parent interest $i$, but node $B$ is in (all of) $i$'s leaf interest(s). Hence, how to deal with this case? Should it be treated as $A$ and $B$ share the same interest? From the perspective of interest-tree, it seems their interests are the same. But from the nodes' perspective, it might be different because of their definitions on the research areas.
\end{itemize}

\begin{table}
\centering
\caption{The number of interests involved each layer of interest-tree before and after emergence}
\label{width}
\begin{tabular}{l l l l l l l l}                                          \hline
Layer No. & 1   & 2   & 3   & 4   & 5   & 6   & 7         \\  \hline
Before emergence & 1     & 4     & 30    & 407   & 116   & 142   & 12          \\ 
After  emergence & 1     & 2     & 19    & 36    & 17    & 7     & 1           \\  \hline
\end{tabular}
\end{table}

\section{Conclusion}
In this paper, we explored the impact of user interests and the relations between them to improve the efficiency of data dissemination in socially aware networking. Specifically, we focused on interest inclusion, one of common relation types between user interests. Inspired by ACM Computing Classification System, we proposed interest-tree to construct our scheme, named Int-Tree. Using this strategy, Int-Tree can identify communities according to interests. Furthermore, Int-Tree takes advantage of density of community and social tie for effective information dissemination and strong adaptability to the dynamic networking environment. The simulation with real dataset proved that Int-Tree outperforms the benchmark protocols of PROPHET and Epidemic with higher delivery ratio, less overhead and less hop counts. The major drawback of Int-Tree lies in that its performance on average latency is not the best over simulation duration or on the circumstances where $Buffer~Size\leq20MB$ or $TTL> 120min$ . Moreover, further simulations on the performance of Int-Tree with changing parameters suggested that weakening the effect of history information can cause a good performance, although $\alpha$, $\beta$ and $\gamma$ do not make dramatic difference.

As a first step on studying relations between user interests, we did not provide experimental evidence to study how multi-interest can affect the dissemination, making the results unsubstantial. However, we put forward some potential challenges and solutions. Meanwhile, in another academic paper, a research has been conducted considering the situation where nodes have multiple interests, accompanied with the intersection of interests. Nevertheless, there are still some problems remained, such as the effect of cross-layer interests and other possible relations. 

\section*{Acknowledgment}
This work is partially supported by the Fundamental Research Funds for the Central Universities (DUT15YQ112) and National Natural Science Foundation of China (61572106).





\bibliographystyle{model1-num-names}
\balance
\bibliography{inttree}

\begin{thebibliography}{42}
\expandafter\ifx\csname natexlab\endcsname\relax\def\natexlab#1{#1}\fi
\providecommand{\url}[1]{\texttt{#1}}
\providecommand{\href}[2]{#2}
\providecommand{\path}[1]{#1}
\providecommand{\DOIprefix}{doi:}
\providecommand{\ArXivprefix}{arXiv:}
\providecommand{\URLprefix}{URL: }
\providecommand{\Pubmedprefix}{pmid:}
\providecommand{\doi}[1]{\href{http://dx.doi.org/#1}{\path{#1}}}
\providecommand{\Pubmed}[1]{\href{pmid:#1}{\path{#1}}}
\providecommand{\bibinfo}[2]{#2}
\ifx\xfnm\relax \def\xfnm[#1]{\unskip,\space#1}\fi
\bibitem[{Pelusi et~al.(2006)Pelusi, Passarella, and
  Conti}]{Pelusi:Opportunistic}
\bibinfo{author}{L.~Pelusi}, \bibinfo{author}{A.~Passarella},
  \bibinfo{author}{M.~Conti},
\newblock \bibinfo{title}{Opportunistic networking: Data forwarding in
  disconnected mobile ad hoc networks},
\newblock \bibinfo{journal}{IEEE Communications Magazine} \bibinfo{volume}{44}
  (\bibinfo{year}{2006}) \bibinfo{pages}{134--141}.
\bibitem[{Warthman(2008)}]{Warthman:DTNs}
\bibinfo{author}{F.~Warthman}, \bibinfo{title}{Delay-tolerant Networks (DTNs) -
  A Tutorial}, \bibinfo{type}{Technical Report}, Warthman Associates,
  \bibinfo{year}{2008}.
\bibitem[{Spyropoulos et~al.(2010)Spyropoulos, Rais, Turletti, Obraczka, and
  Vasilakos}]{Spyropoulos:Routing}
\bibinfo{author}{T.~Spyropoulos}, \bibinfo{author}{R.~N.~B. Rais},
  \bibinfo{author}{T.~Turletti}, \bibinfo{author}{K.~Obraczka},
  \bibinfo{author}{A.~Vasilakos},
\newblock \bibinfo{title}{Routing for disruption tolerant networks: Taxonomy
  and design},
\newblock \bibinfo{journal}{Wireless Networks} \bibinfo{volume}{16}
  (\bibinfo{year}{2010}) \bibinfo{pages}{2349--2370}.
\bibitem[{Xia et~al.(2015)Xia, Liu, Li, Ma, and Vasilakos}]{Xia:SAN}
\bibinfo{author}{F.~Xia}, \bibinfo{author}{L.~Liu}, \bibinfo{author}{J.~Li},
  \bibinfo{author}{J.~Ma}, \bibinfo{author}{A.~V. Vasilakos},
\newblock \bibinfo{title}{Socially aware networking: A survey},
\newblock \bibinfo{journal}{IEEE Systems Journal} \bibinfo{volume}{9}
  (\bibinfo{year}{2015}) \bibinfo{pages}{904--921}.
\bibitem[{Vastardis and Yang(2013)}]{Vastardis:Multi-phase}
\bibinfo{author}{N.~Vastardis}, \bibinfo{author}{K.~Yang},
\newblock \bibinfo{title}{Multi-phase socially-aware routing in distributed
  mobile social networks},
\newblock in: \bibinfo{booktitle}{Proceedings of 2013 9th International
  Wireless Communications and Mobile Computing Conference (IEEE IWCMC 2013)},
  \bibinfo{address}{Cagliari, Sardinia Italy}, \bibinfo{year}{2013}, pp.
  \bibinfo{pages}{1353--1358}.
\bibitem[{Zhu et~al.(2013)Zhu, Xu, Shi, and Wang}]{RoutingSurvey-Zhu}
\bibinfo{author}{Y.~Zhu}, \bibinfo{author}{B.~Xu}, \bibinfo{author}{X.~Shi},
  \bibinfo{author}{Y.~Wang},
\newblock \bibinfo{title}{A survey of social-based routing in delay tolerant
  networks: Positive and negative social effects},
\newblock \bibinfo{journal}{IEEE Communications and Surveys Tutorials}
  \bibinfo{volume}{15} (\bibinfo{year}{2013}) \bibinfo{pages}{387--401}.
\bibitem[{Xia et~al.(2014)Xia, Ahmed, Yang, Ma, and Rodrigues}]{Xia-TPDS2014}
\bibinfo{author}{F.~Xia}, \bibinfo{author}{A.~M. Ahmed}, \bibinfo{author}{L.~T.
  Yang}, \bibinfo{author}{J.~Ma}, \bibinfo{author}{J.~Rodrigues},
\newblock \bibinfo{title}{Exploiting social relationship to enable efficient
  replica allocation in ad-hoc social networks},
\newblock \bibinfo{journal}{IEEE Transactions on Parallel and Distributed
  Systems} \bibinfo{volume}{25} (\bibinfo{year}{2014})
  \bibinfo{pages}{3167--3176}.
\bibitem[{Xia et~al.(2015)Xia, Ahmed, Yang, and Luo}]{Xia-TC2014}
\bibinfo{author}{F.~Xia}, \bibinfo{author}{A.~M. Ahmed}, \bibinfo{author}{L.~T.
  Yang}, \bibinfo{author}{Z.~Luo},
\newblock \bibinfo{title}{Community-based event dissemination with optimal load
  balancing},
\newblock \bibinfo{journal}{IEEE Transactions on Computers}
  \bibinfo{volume}{64} (\bibinfo{year}{2015}) \bibinfo{pages}{1857--1869}.
\bibitem[{Wasserman and Faust(1994)}]{SNA-book}
\bibinfo{author}{S.~Wasserman}, \bibinfo{author}{K.~Faust},
  \bibinfo{title}{Social network analysis: methods and applications, structural
  analysis in the social sciences series}, \bibinfo{publisher}{Cambridge Univ.
  Press}, \bibinfo{year}{1994}.
\bibitem[{Fan et~al.(2013)Fan, Chen, Du, Gao, Wu, and Sun}]{Fan:Geocommunity}
\bibinfo{author}{J.~Fan}, \bibinfo{author}{J.~Chen}, \bibinfo{author}{Y.~Du},
  \bibinfo{author}{W.~Gao}, \bibinfo{author}{J.~Wu}, \bibinfo{author}{Y.~Sun},
\newblock \bibinfo{title}{Geocommunity-based broadcasting for data
  dissemination in mobile social networks},
\newblock \bibinfo{journal}{IEEE Transactions on Parallel and Distributed
  Systems} \bibinfo{volume}{24} (\bibinfo{year}{2013})
  \bibinfo{pages}{734--743}.
\bibitem[{Xiao et~al.(2014)Xiao, Wu, and Huang}]{Xiao:Community-aware}
\bibinfo{author}{M.~Xiao}, \bibinfo{author}{J.~Wu}, \bibinfo{author}{L.~Huang},
\newblock \bibinfo{title}{Community-aware opportunistic routing in mobile
  social networks},
\newblock \bibinfo{journal}{IEEE Transactions on Computers}
  \bibinfo{volume}{63} (\bibinfo{year}{2014}) \bibinfo{pages}{1682--1695}.
\bibitem[{Wu et~al.(2013)Wu, Xiao, and Huang}]{Wu:Homing}
\bibinfo{author}{J.~Wu}, \bibinfo{author}{M.~Xiao}, \bibinfo{author}{L.~Huang},
\newblock \bibinfo{title}{Homing spread: Community home-based multi-copy
  routing in mobile social networks},
\newblock in: \bibinfo{booktitle}{Proceedings of the 32nd IEEE International
  Conference on Computer Communciations (IEEE INFOCOM 2013)},
  \bibinfo{year}{2013}, pp. \bibinfo{pages}{2319--2327}.
\bibitem[{McPherson et~al.(2001)McPherson, Smith-Lovin, and
  Cook}]{McPherson:Homophily}
\bibinfo{author}{M.~McPherson}, \bibinfo{author}{L.~Smith-Lovin},
  \bibinfo{author}{J.~M. Cook},
\newblock \bibinfo{title}{Birds of a feather: Homophily in social networks},
\newblock \bibinfo{journal}{Annual Review of Sociology} \bibinfo{volume}{27}
  (\bibinfo{year}{2001}) \bibinfo{pages}{415--444}.
\bibitem[{Hidi(2006)}]{Hidi:Interest}
\bibinfo{author}{S.~Hidi},
\newblock \bibinfo{title}{Interest: A unique motivational variable},
\newblock \bibinfo{journal}{Educational Research Review} \bibinfo{volume}{1}
  (\bibinfo{year}{2006}) \bibinfo{pages}{69--82}.
\bibitem[{Zhou et~al.(2008)Zhou, Kiet, Kim, Wang, and Holme}]{Zhou:Role}
\bibinfo{author}{T.~Zhou}, \bibinfo{author}{H.~A.~T. Kiet},
  \bibinfo{author}{B.~J. Kim}, \bibinfo{author}{B.-H. Wang},
  \bibinfo{author}{P.~Holme},
\newblock \bibinfo{title}{Role of activity in human dynamics},
\newblock \bibinfo{journal}{EPL (Europhysics Letters)} \bibinfo{volume}{82}
  (\bibinfo{year}{2008}) \bibinfo{pages}{28002}.
\bibitem[{Han et~al.(2008)Han, Zhou, and Wang}]{Han:Modeling}
\bibinfo{author}{X.-P. Han}, \bibinfo{author}{T.~Zhou}, \bibinfo{author}{B.-H.
  Wang},
\newblock \bibinfo{title}{Modeling human dynamics with adaptive interest},
\newblock \bibinfo{journal}{New Journal of Physics} \bibinfo{volume}{10}
  (\bibinfo{year}{2008}) \bibinfo{pages}{073010}.
\bibitem[{Shang et~al.(2006)Shang, Chen, Dai, Wang, and
  Zhou}]{Shang:Interest-Driven}
\bibinfo{author}{M.-S. Shang}, \bibinfo{author}{G.-X. Chen},
  \bibinfo{author}{S.-X. Dai}, \bibinfo{author}{B.-H. Wang},
  \bibinfo{author}{T.~Zhou},
\newblock \bibinfo{title}{Interest-driven model for human dynamics},
\newblock \bibinfo{journal}{Chinese Physics Letters} \bibinfo{volume}{27}
  (\bibinfo{year}{2006}) \bibinfo{pages}{048701}.
\bibitem[{Chen et~al.(2014)Chen, Shen, and Zhang}]{Chen:Leveraging}
\bibinfo{author}{K.~Chen}, \bibinfo{author}{H.~Shen},
  \bibinfo{author}{H.~Zhang},
\newblock \bibinfo{title}{Leveraging social networks for p2p content-based file
  sharing in disconnected manets},
\newblock \bibinfo{journal}{IEEE Transactions on Mobile Computing}
  \bibinfo{volume}{13} (\bibinfo{year}{2014}) \bibinfo{pages}{235--249}.
\bibitem[{Costa et~al.(2008)Costa, Mascolo, Musolesi, and
  Picco}]{Costa:Socially-aware}
\bibinfo{author}{P.~Costa}, \bibinfo{author}{C.~Mascolo},
  \bibinfo{author}{M.~Musolesi}, \bibinfo{author}{G.~P. Picco},
\newblock \bibinfo{title}{Socially-aware routing for publish-subscribe in
  delay-tolerant mobile ad hoc networks},
\newblock \bibinfo{journal}{IEEE Journal on Selected Areas in Communciations
  Special issue on Delay-Tolerant Networks} \bibinfo{volume}{26}
  (\bibinfo{year}{2008}) \bibinfo{pages}{748--760}.
\bibitem[{Zhu et~al.(2011)Zhu, Wang, and Wang}]{Zhu:Ripple}
\bibinfo{author}{Y.~Zhu}, \bibinfo{author}{J.~Wang}, \bibinfo{author}{C.~Wang},
\newblock \bibinfo{title}{Ripple: A publish/subscribe service for multidata
  item updates propagation in the cloud},
\newblock \bibinfo{journal}{Journal of Network and Computer Applications}
  \bibinfo{volume}{34} (\bibinfo{year}{2011}) \bibinfo{pages}{1054--1067}.
\bibitem[{Chuah et~al.(2010)Chuah, Yang, and Hui}]{Chuah:Cooperative}
\bibinfo{author}{M.~Chuah}, \bibinfo{author}{P.~Yang},
  \bibinfo{author}{P.~Hui},
\newblock \bibinfo{title}{Cooperative user centric information dissemination in
  human content-based networks},
\newblock in: \bibinfo{booktitle}{Proceedings of th 16th IEEE International
  Conference on Parallel and Distributed Systems (IEEE ICPADS 2010)},
  \bibinfo{address}{Nagoya, Japan}, \bibinfo{year}{2010}, pp.
  \bibinfo{pages}{794--799}.
\bibitem[{Xia et~al.(2015)Xia, Liu, Li, Ahmed, Yang, and Ma}]{Xia:BEEINFO}
\bibinfo{author}{F.~Xia}, \bibinfo{author}{L.~Liu}, \bibinfo{author}{J.~Li},
  \bibinfo{author}{A.~M. Ahmed}, \bibinfo{author}{L.~T. Yang},
  \bibinfo{author}{J.~Ma},
\newblock \bibinfo{title}{Beeinfo: An interest-based forwarding scheme using
  artificial bee colony for socially-aware networking},
\newblock \bibinfo{journal}{IEEE Transactions on Vehicular Technology}
  \bibinfo{volume}{64} (\bibinfo{year}{2015}) \bibinfo{pages}{1188--1200}.
\bibitem[{Vahdat and Becker(2000)}]{Vahat:Epidemic}
\bibinfo{author}{A.~Vahdat}, \bibinfo{author}{D.~Becker},
  \bibinfo{title}{Epidemic routing for partially connected ad hoc networks},
  \bibinfo{type}{Technical Report}, Technical Report CS-200006, Duke
  University, \bibinfo{year}{2000}.
\bibitem[{Lindgren et~al.(2004)Lindgren, Doria, and
  Schel\'{e}n}]{Lindgren:Probabilistic}
\bibinfo{author}{A.~Lindgren}, \bibinfo{author}{A.~Doria},
  \bibinfo{author}{O.~Schel\'{e}n},
\newblock \bibinfo{title}{Probabilistic routing in intermittently connected
  networks},
\newblock in: \bibinfo{editor}{P.~Dini}, \bibinfo{editor}{P.~Lorenz},
  \bibinfo{editor}{J.~N. de~Souza} (Eds.), \bibinfo{booktitle}{Service
  Assurance with Partial and Intermittent Resources}, volume
  \bibinfo{volume}{3126} of \textit{\bibinfo{series}{Lecture Notes in Computer
  Science}}, \bibinfo{publisher}{Springer Berlin Heidelberg},
  \bibinfo{year}{2004}, pp. \bibinfo{pages}{239--254}.
\bibitem[{Yan et~al.(2012)Yan, Wu, and Yi}]{Yan:Modeling}
\bibinfo{author}{Q.~Yan}, \bibinfo{author}{L.~Wu}, \bibinfo{author}{L.~Yi},
\newblock \bibinfo{title}{Modeling of posting behavior in mobile internet based
  on human dynamics},
\newblock in: \bibinfo{booktitle}{Proceedings of 2012 Second International
  Conference Cloud and Green Computing (IEEE CGC 2012)},
  \bibinfo{address}{Hunan, China}, \bibinfo{year}{2012}, pp.
  \bibinfo{pages}{744--747}.
\bibitem[{Carofiglio et~al.(2012)Carofiglio, Gallo, and
  Muscariello}]{Carofiglio:ICP}
\bibinfo{author}{G.~Carofiglio}, \bibinfo{author}{M.~Gallo},
  \bibinfo{author}{L.~Muscariello},
\newblock \bibinfo{title}{Icp: Design and evaluation of an interest control
  protocol for content-centric networking},
\newblock in: \bibinfo{booktitle}{Proceedings of the 31st Annual IEEE
  International Conference on Computer Communications Workshops (IEEE INFOCOM
  2012)}, \bibinfo{address}{Orlando, Florida USA}, \bibinfo{year}{2012}, pp.
  \bibinfo{pages}{304--309}.
\bibitem[{Bjurefors et~al.(2010)Bjurefors, Gunningberg, Nordstr{\"o}m, and
  Rohner}]{Bjurefors:Interest}
\bibinfo{author}{F.~Bjurefors}, \bibinfo{author}{P.~Gunningberg},
  \bibinfo{author}{E.~Nordstr{\"o}m}, \bibinfo{author}{C.~Rohner},
\newblock \bibinfo{title}{Interest dissemination in a searchable data-centric
  opportunistic network},
\newblock in: \bibinfo{booktitle}{Proceedings of 2010 European Wireless
  Conference}, \bibinfo{address}{Lucca, Tuscany Italy}, \bibinfo{year}{2010},
  pp. \bibinfo{pages}{889--895}.
\bibitem[{Gao and Cao(2011)}]{Gao:User}
\bibinfo{author}{W.~Gao}, \bibinfo{author}{G.~Cao},
\newblock \bibinfo{title}{User-centric data dissemination in disruption
  tolerant networks},
\newblock in: \bibinfo{booktitle}{Proceedings of the 30th IEEE International
  Conference on Computer Communications (IEEE INFOCOM 2011)},
  \bibinfo{address}{Shanghai, China}, \bibinfo{year}{2011}, pp.
  \bibinfo{pages}{3119--3127}.
\bibitem[{Wu and Wang(2014)}]{Wu:Hypercube}
\bibinfo{author}{J.~Wu}, \bibinfo{author}{Y.~Wang},
\newblock \bibinfo{title}{Hypercube-based multipath social feature routing in
  human contact networks},
\newblock \bibinfo{journal}{IEEE Transactions on Computers}
  \bibinfo{volume}{63} (\bibinfo{year}{2014}) \bibinfo{pages}{383--396}.
\bibitem[{Hui and Crowcroft(2007)}]{Hui:How}
\bibinfo{author}{P.~Hui}, \bibinfo{author}{J.~Crowcroft},
\newblock \bibinfo{title}{How small labels create big improvements},
\newblock in: \bibinfo{booktitle}{Proceedings of the 5th Annual IEEE
  International Conference on Pervasive Computing and Communications Workshops
  (IEEE PerCom 2007)}, \bibinfo{address}{New York, USA}, \bibinfo{year}{2007},
  pp. \bibinfo{pages}{65--70}.
\bibitem[{Hui et~al.(2009)Hui, Crowcroft, and Yoneki}]{Hui:Bubblerap}
\bibinfo{author}{P.~Hui}, \bibinfo{author}{J.~Crowcroft},
  \bibinfo{author}{E.~Yoneki},
\newblock \bibinfo{title}{Bubble rap: Social -based forwarding in delay
  tolerant networks},
\newblock in: \bibinfo{booktitle}{Proceedings of ACM International Symposiumo
  on Mobile Ad Hoc Networking and Computing (ACM MobiHoc 2008)},
  \bibinfo{address}{Hong Kong}, \bibinfo{year}{2009}, pp.
  \bibinfo{pages}{1--9}.
\bibitem[{Li and Wu(2009)}]{Li:Localcom}
\bibinfo{author}{F.~Li}, \bibinfo{author}{J.~Wu},
\newblock \bibinfo{title}{Localcom: A community-based epidemic forwarding
  scheme in disruption-tolerant networks},
\newblock in: \bibinfo{booktitle}{Proceedings of IEEE International Conference
  on Sensor, Mesh and Ad Hoc Communications and Networks (IEEE SECON 2009)},
  \bibinfo{address}{Rome, Italy}, \bibinfo{year}{2009}, pp.
  \bibinfo{pages}{1--9}.
\bibitem[{Musolesi et~al.(2008)Musolesi, Hui, Mascolo, and
  Crowcroft}]{Musolesi:Writing}
\bibinfo{author}{M.~Musolesi}, \bibinfo{author}{P.~Hui},
  \bibinfo{author}{C.~Mascolo}, \bibinfo{author}{J.~Crowcroft},
\newblock \bibinfo{title}{Writing on the clean slate: Implementing a
  socially-aware protocol in haggle},
\newblock in: \bibinfo{booktitle}{Proceedings of IEEE International Symposium
  on a World of Wireless Mobile and Multimedia Networks (IEEE WoWMoM 2008)},
  \bibinfo{address}{Newport Beach, USA}, \bibinfo{year}{2008}, pp.
  \bibinfo{pages}{1--6}.
\bibitem[{Musolesi et~al.(2005)Musolesi, Hailes, and
  Mascolo}]{Musolesi:Adaptive}
\bibinfo{author}{M.~Musolesi}, \bibinfo{author}{S.~Hailes},
  \bibinfo{author}{C.~Mascolo},
\newblock \bibinfo{title}{Adaptive routing for intermittently connected mobile
  ad hoc networks},
\newblock in: \bibinfo{booktitle}{Proceedings of IEEE International Symposium
  on a World of Wireless Mobile and Multimedia Networks (IEEE WoWMoM 2005)},
  \bibinfo{address}{Taormina - Giardini Naxos, Italy}, \bibinfo{year}{2005},
  pp. \bibinfo{pages}{183--189}.
\bibitem[{Katsaros et~al.(2010)Katsaros, Dimokas, and
  Tassiulas}]{Katsaros:Social}
\bibinfo{author}{D.~Katsaros}, \bibinfo{author}{N.~Dimokas},
  \bibinfo{author}{L.~Tassiulas},
\newblock \bibinfo{title}{Social network analysis concepts in the design of
  wireless ad hoc network protocols},
\newblock \bibinfo{journal}{IEEE Network} \bibinfo{volume}{24}
  (\bibinfo{year}{2010}) \bibinfo{pages}{23--29}.
\bibitem[{Fast et~al.(2005)Fast, Jensen, and Levine}]{Fast:Creating}
\bibinfo{author}{A.~Fast}, \bibinfo{author}{D.~Jensen}, \bibinfo{author}{B.~N.
  Levine},
\newblock \bibinfo{title}{Creating social networks to improve peer-to-peer
  networking},
\newblock in: \bibinfo{booktitle}{Proceedings of the 11th ACM SIGKDD
  International Conference on Knowledge Discovery in Data Mining (ACM KDD
  2005)}, \bibinfo{address}{Chicago, USA}, \bibinfo{year}{2005}, pp.
  \bibinfo{pages}{568--573}.
\bibitem[{Ruiz et~al.(2012)Ruiz, Dorronsoro, Bouvry, and
  Tard{\'o}n}]{Ruiz:Information}
\bibinfo{author}{P.~Ruiz}, \bibinfo{author}{B.~Dorronsoro},
  \bibinfo{author}{P.~Bouvry}, \bibinfo{author}{L.~Tard{\'o}n},
\newblock \bibinfo{title}{Information dissemination in vanets based upon a tree
  topology},
\newblock \bibinfo{journal}{Ad Hoc Networks} \bibinfo{volume}{10}
  (\bibinfo{year}{2012}) \bibinfo{pages}{111--127}.
\bibitem[{Jain et~al.(2011)Jain, Ramakrishnan, and Chiu}]{Vendramin:GrAnt}
\bibinfo{author}{R.~Jain}, \bibinfo{author}{K.~K. Ramakrishnan},
  \bibinfo{author}{D.~M. Chiu}, \bibinfo{title}{GrAnt: Inferring Best
  Forwarders from Complex Networks' Dynamics through a Greedy Ant Colony
  Optimization}, \bibinfo{type}{Technical Report} \bibinfo{number}{RR-7694},
  INRIA, \bibinfo{year}{2011}.
\bibitem[{Ioannidis and Chaintreau(2009)}]{Ioannidis:Strength}
\bibinfo{author}{S.~Ioannidis}, \bibinfo{author}{A.~Chaintreau},
\newblock \bibinfo{title}{On the strength of weak ties in mobile social
  networks},
\newblock in: \bibinfo{booktitle}{Proceedings of the Second ACM EuroSys
  Workshop on Social Network Systems (ACM SNS 2009)},
  \bibinfo{address}{Nuremberg, Germany}, \bibinfo{year}{2009}, pp.
  \bibinfo{pages}{19--25}.
\bibitem[{Pietil\"{a}inen et~al.(2009)Pietil\"{a}inen, Oliver, LeBrun,
  Varghese, and Diot}]{SIGCOMM2009}
\bibinfo{author}{A.-K. Pietil\"{a}inen}, \bibinfo{author}{E.~Oliver},
  \bibinfo{author}{J.~LeBrun}, \bibinfo{author}{G.~Varghese},
  \bibinfo{author}{C.~Diot},
\newblock \bibinfo{title}{Mobiclique: Middleware for mobile social networking},
\newblock in: \bibinfo{booktitle}{Proceedings of the 2nd ACM Workshop on Online
  Social Networks (ACM WOSN 2009)}, \bibinfo{address}{Barcelona, Spain},
  \bibinfo{year}{2009}, pp. \bibinfo{pages}{49--54}.
\bibitem[{Ker{\"a}nen et~al.(2010)Ker{\"a}nen, K{\"a}rkk{\"a}inen, and
  Ott}]{Keranen:Simulating}
\bibinfo{author}{A.~Ker{\"a}nen}, \bibinfo{author}{T.~K{\"a}rkk{\"a}inen},
  \bibinfo{author}{J.~Ott},
\newblock \bibinfo{title}{Simulating mobility and dtns with the one (invited
  paper)},
\newblock \bibinfo{journal}{Journal of Communications} \bibinfo{volume}{5}
  (\bibinfo{year}{2010}) \bibinfo{pages}{92--105}.
\bibitem[{Chuang et~al.(2013)Chuang, Lin, and Chen}]{Chuang:Bluetooth}
\bibinfo{author}{K.-T. Chuang}, \bibinfo{author}{Y.-J. Lin},
  \bibinfo{author}{C.-C. Chen},
\newblock \bibinfo{title}{Bluetooth-based mobile p2p framework for
  preference-aware data dissemination on social networks},
\newblock in: \bibinfo{booktitle}{Proceedings of 2013 IEEE 14th International
  Conference on Mobile Data Management (IEEE MDM 2013)},
  volume~\bibinfo{volume}{2}, \bibinfo{address}{Milan, Italy},
  \bibinfo{year}{2013}, pp. \bibinfo{pages}{110--115}.

\end{thebibliography}







\end{document}